\documentclass[twocolumn]{emulateapj}
\usepackage{graphicx}
\usepackage{amssymb}
\usepackage{amsmath}
\usepackage{subfigure}
\usepackage{multirow}
\usepackage{threeparttable}
\usepackage{array}
\usepackage{natbib}
\usepackage{xcolor}
\usepackage{subfigure}
\usepackage{hyperref}

\hypersetup{
    colorlinks=true,
    linkcolor=red,
    filecolor=magenta,
    urlcolor=cyan,
}

\def\kms{~\rm km~s^{-1}}
\def\cmsq{~\rm cm^{-2}}
\def\cc{~\rm cm^{-3}}

\defcitealias{Qu:2020aa}{Paper I}

\begin{document}
\title{The Warm Gas in the MW: the Kinematical Model of C IV and its Connection with Si IV}

\author{Zhijie Qu}
\affiliation{Department of Astronomy, University of Michigan, Ann Arbor, MI 48109, USA}

\author{Ryan Lindley}
\affiliation{Department of Astronomy, University of Michigan, Ann Arbor, MI 48109, USA}

\author{Joel N. Bregman}
\affiliation{Department of Astronomy, University of Michigan, Ann Arbor, MI 48109, USA}

\email{quzhijie@umich.edu}

\begin{abstract}
We compose a 265-sight line MW \ion{C}{4} line shape sample using the {\it HST}/COS archive, which is complementary to the existing \ion{Si}{4} samples.
\ion{C}{4} has a higher ionization potential ($47 - 64$ eV) than \ion{Si}{4} ($33 - 45$ eV), so it also traces warm gas, which is roughly cospatial with \ion{Si}{4}.
The spatial density distribution and kinematics of \ion{C}{4} is identical with \ion{Si}{4} within $\approx 2 \sigma$.
\ion{C}{4} is more sensitive to the warm gas density distribution at large radii with a higher element abundance.
Applying the kinematical model to the \ion{C}{4} sample, we find two possible solutions of the density distribution, which are distinguished by the relative extension along the disk mid-plane and the normal-line direction.
Both two solutions can reproduce the existing sample, and suggest a warm gas disk mass of $\log M(M_\odot) \approx 8$ and an upper limit of $\log M(M_\odot) < 9.3$ within 250 kpc, which is consistent with \ion{Si}{4}.
There is a decrease of the \ion{C}{4}/\ion{Si}{4} column density ratio from the Galactic center (GC) to outskirts by $0.2-0.3$ dex, which may suggest a phase transition or different ionization mechanisms for \ion{C}{4} and \ion{Si}{4}.
Also, we find that the difference between \ion{C}{4} and \ion{Si}{4} is an excellent tracer of small-scale features, and we find a typical size of $5^\circ-10^\circ$ for possible turbulence within individual clouds ($\approx 1\rm~kpc$).
\end{abstract}

\section{Introduction}

The circumgalactic medium (CGM) is a baryon reservoir surrounding the galaxy disk, which is deeply involved in the baryonic cycle of galaxy evolution (see \citealt{Tumlinson:2017aa} for a review).
In the past two decades, extensive observational studies revealed that the CGM is massive, accounting for $30-50\%$ of the total baryons in an $L^*$ galaxy \citep[e.g.][]{Werk:2014aa, Miller:2015aa, Prochaska:2017aa, Bregman:2018aa}.
The CGM is composed of a multi-phase medium spanning a temperature range from the  the cold gas ($T\approx 10$ K) to the hot gas ($T \gtrsim 10^{6}$ K; e.g., \citealt{Menard:2010aa, Li:2013aa, Stocke:2013aa, Fang:2015aa, Peek:2015aa, Bogdan:2017aa, Burchett:2019aa, Das:2019aa, Lehner:2020aa, Kaaret:2020aa, Bregman:2021aa}).
This massive multi-phase CGM not only supplies the gaseous disk for continuous star formation, but also gathers the feedback materials from stellar evolution.
Therefore, there is a gas exchange (i.e., interaction) between the disk and the CGM (i.e., the disk-CGM interaction zone; \citealt{Savage:2009aa, Zheng:2017aa, Werk:2019aa}), which is crucial to understanding galaxy evolution.

Among the multi-phase CGM, the warm gas ($T \approx 10^{5}$ K) is of great importance, because it is the most active phase due to its short lifetime (i.e., rapid radiative cooling $\approx 10$ Myr; \citealt{Oppenheimer:2013aa, Gnat:2017aa}).
The short lifetime of the warm gas suggests it must be refreshed frequently by interaction processes.
Particularly in the CGM, the warm gas is generated by accretion and feedback processes \citep[e.g.,][]{Kim:2018aa, Stern:2018aa, Li:2020aa}, so it is an excellent tracer of these processes.

The warm gas is traced by intermediate ionization state ions, such as \ion{C}{4}, \ion{Si}{4}, and \ion{O}{6}, which have resonant lines in the UV band.
In the local universe, these ions are observed as absorption lines against background UV-bright targets (hot stars or active galactic nuclei; AGN).
Since the UV-bright sources are rare, the Milky Way (MW) is the only galaxy with hundreds of sight lines to map the CGM towards all directions.
For the MW, the warm gas distribution and kinematics are studied by measuring the column densities and velocities in the past decades \citep[e.g.,][]{Jenkins:1978aa, Savage:1981aa, de-Boer:1983aa, Sembach:1992aa,  Shull:1994aa, Savage:2003aa, Sembach:2003aa, Fox:2004aa, Bowen:2008aa, Savage:2009aa, Lehner:2011aa, Wakker:2012aa, Fox:2014aa, Fox:2019aa, Werk:2019aa, Zheng:2019aa, Qu:2019aa, Bish:2021aa}.
These studies suggested that the MW has a thick warm gas disk (e.g., \citealt{Savage:2009aa}) and a massive warm gas CGM (e.g., \citealt{Zheng:2019aa}).
However, the warm gas is still poorly constrained at large radii because of the viewing geometry from the disk.

In \citet[][hereafter \citetalias{Qu:2020aa}]{Qu:2020aa}, we extracted an all-sky MW \ion{Si}{4} column density line shape sample, and developed a kinematical model by combining the density distribution and kinematics to reproduce observed line shapes.
In this model, the density distribution is approximated by an exponential disk and a power-law CGM component, while the kinematics is approximated by both rotation and radial motions.
We applied this kinematical model to the \ion{Si}{4} sample, and found the large scale features of the \ion{Si}{4} gas can be modeled as a thick disk co-rotating with the stellar disk.
After subtracting the large-scale model, we also found small-scale variations in the residual maps for both column density and velocity shifts, which preferred a galactic fountain scenario for the origin of the warm gas in the MW.

Here, we extend this study to another intermediate ionization state ion \ion{C}{4}, which is complementary to the \ion{Si}{4} sample, because \ion{C}{4} ($47 -64$ eV) has a higher potential than \ion{Si}{4} ($33- 45$ eV) and is a more abundant element.
Also, we extend the selection criterion to a lower $S/N = 5$ to construct a large joint sample with both \ion{Si}{4} and \ion{C}{4}.
At large scales, these two ions are roughly cospatial, so the inclusion of the \ion{C}{4} sample can improve the significance of kinematical modeling.
At small scales, the difference between \ion{C}{4} and \ion{Si}{4} allows one to study small scale structure without  model dependences (different from the kinematical model correction; \citetalias{Qu:2020aa}).

In this paper, the sample and data reduction are introduced in Section 2.
The kinematical model developed in \citetalias{Qu:2020aa} is briefly described and applied to the \ion{C}{4} sample in Section 3.
In Section 4, we report the all-sky variation of the \ion{Si}{4} and \ion{C}{4} difference as the ion ratio and the velocity shift.
The key results are summarized in Section 5.

\section{Sample and Data Reduction}
For the MW, there are two types of sight lines -- the stellar sight lines within (or close to) the MW disk and the distant AGN sight lines.
The AGN sight lines are employed to trace gas at large radii, and we extract the line shape sample of AGN sight lines in the {\it Hubble Space Telescope} ({\it HST}) Spectroscopic Legacy Archive (HSLA; \citealt{Peeples:2017aa}).
The MW stellar sight lines only trace the nearby gas, but it is also needed to constrain the foreground gas distribution, and decompose gas at large radii (i.e., the CGM and the disk-CGM interaction zone).
The construction of the stellar sight line sample is the same as \citetalias{Qu:2020aa}.
We adopt stellar sight lines from the \citet{Savage:2009aa} and \citet{Lehner:2011aa} surveys. 
For the \ion{C}{4} stellar sample, there are 50 measurements, 5 lower limits, and 7 upper limits from the {\it IUE} sample \citep{Savage:2009aa}, and another 27 measurements from the {\it HST/STIS} sample \citep{Lehner:2011aa}.
In the {\it IUE} sample, sight lines are omitted if they were also observed by {\it HST/STIS}.

The AGN sight line sample is extracted from the HSLA archive with a limiting continuum signal-to-noise ratio of $S/N \geq 5$.
For the \ion{C}{4} sample, there are 265 sights lights with $S/N > 5$, among which 151 sight lines have $S/N > 10$.
As argued in \citetalias{Qu:2020aa}, the kinematical model fitting is dominated by the high $S/N > 10$ sample.
In this work, we also investigate the sight line variation for the \ion{C}{4} and \ion{Si}{4} difference, so we extend the threshold to $S/N>5$, which increases the joint sample by a factor of 2.
In total, there are 203 matched \ion{Si}{4} sight lines with $S/N>5$, among which 112 sight lines have \ion{Si}{4} and \ion{C}{4} with $S/N > 10$.

In \citetalias{Qu:2020aa}, we used the apparent optical depth method (AODM) to extract the column density line profile \citep{Savage:1991aa}.
This method could only be applied to the linear region of the curve of growth, because saturation leads to an underestimation of the peak column density \citep{Savage:1991aa, Jenkins:1996aa}.
As stated in \citetalias{Qu:2020aa}, only a few \ion{Si}{4} sight lines are affected by saturation, and only in the strong line of the doublet.
Therefore, we adopt the AODM for the \ion{Si}{4} reduction in \citetalias{Qu:2020aa}.
However, because carbon has a higher abundance than silicon, about half of \ion{C}{4} absorption features are affected by saturation for both strong and weak lines.
Here, we introduce a hybrid method combining the AODM and Voigt profile fitting for the \ion{C}{4} line shape extraction.

The basic idea is to use the Voigt profile to define the first-order approximation of the line shape, which accounts for saturation.
The hybrid method works for the moderate saturation, which is similar to the AODM correction based on the doublet flux ratio \citep{Jenkins:1996aa}.
Our correction is different from the pixel-by-pixel correction in \citet{Jenkins:1996aa} in that it is applied to individual absorption components, so it can be used in lower $S/N $ systems ($5 < S/N \lesssim 10$).
The advantage of the new method is the small uncertainty for the correction factor. 

\begin{figure*}
\begin{center}
\includegraphics[width=0.49\textwidth]{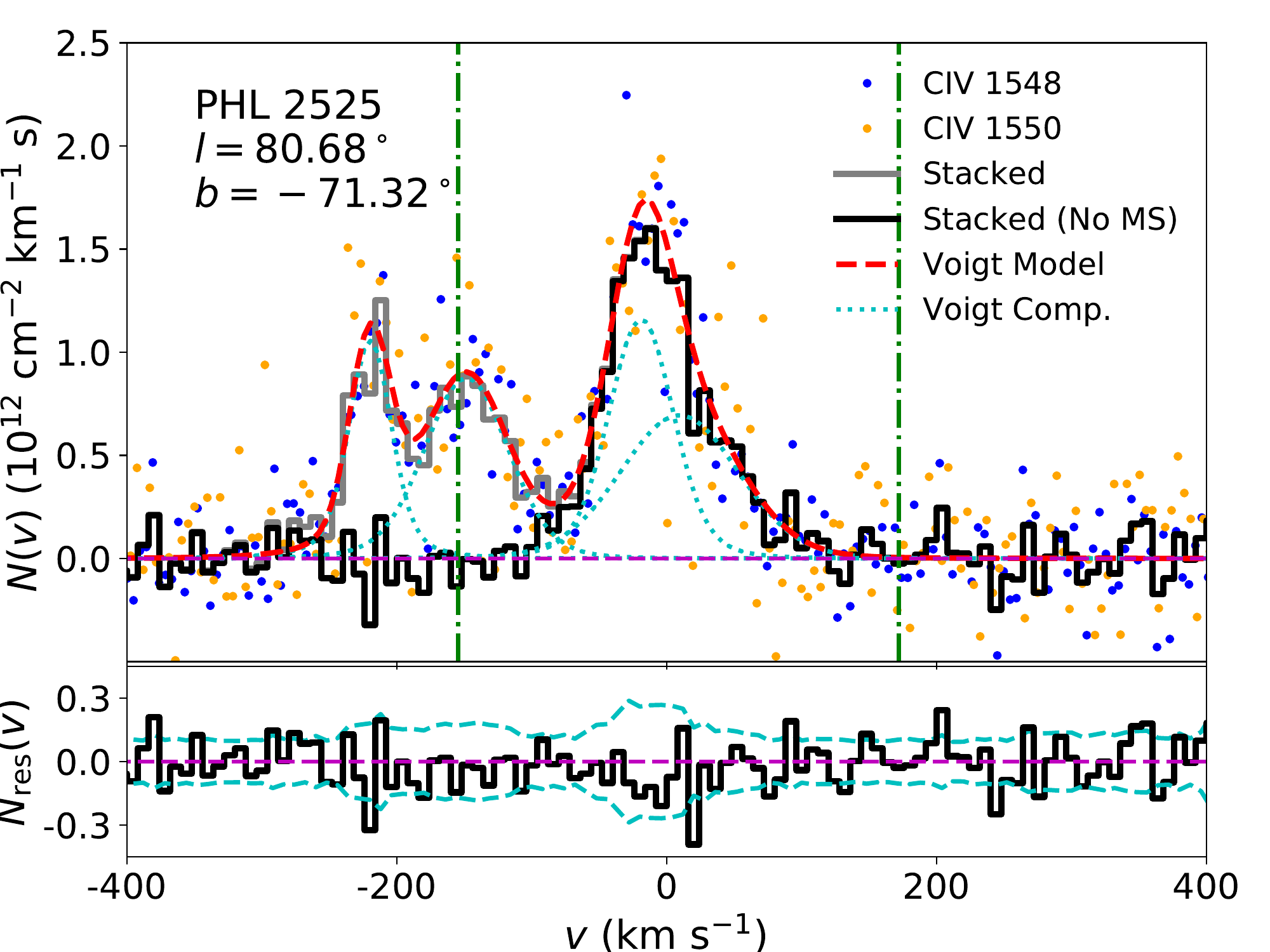}
\includegraphics[width=0.49\textwidth]{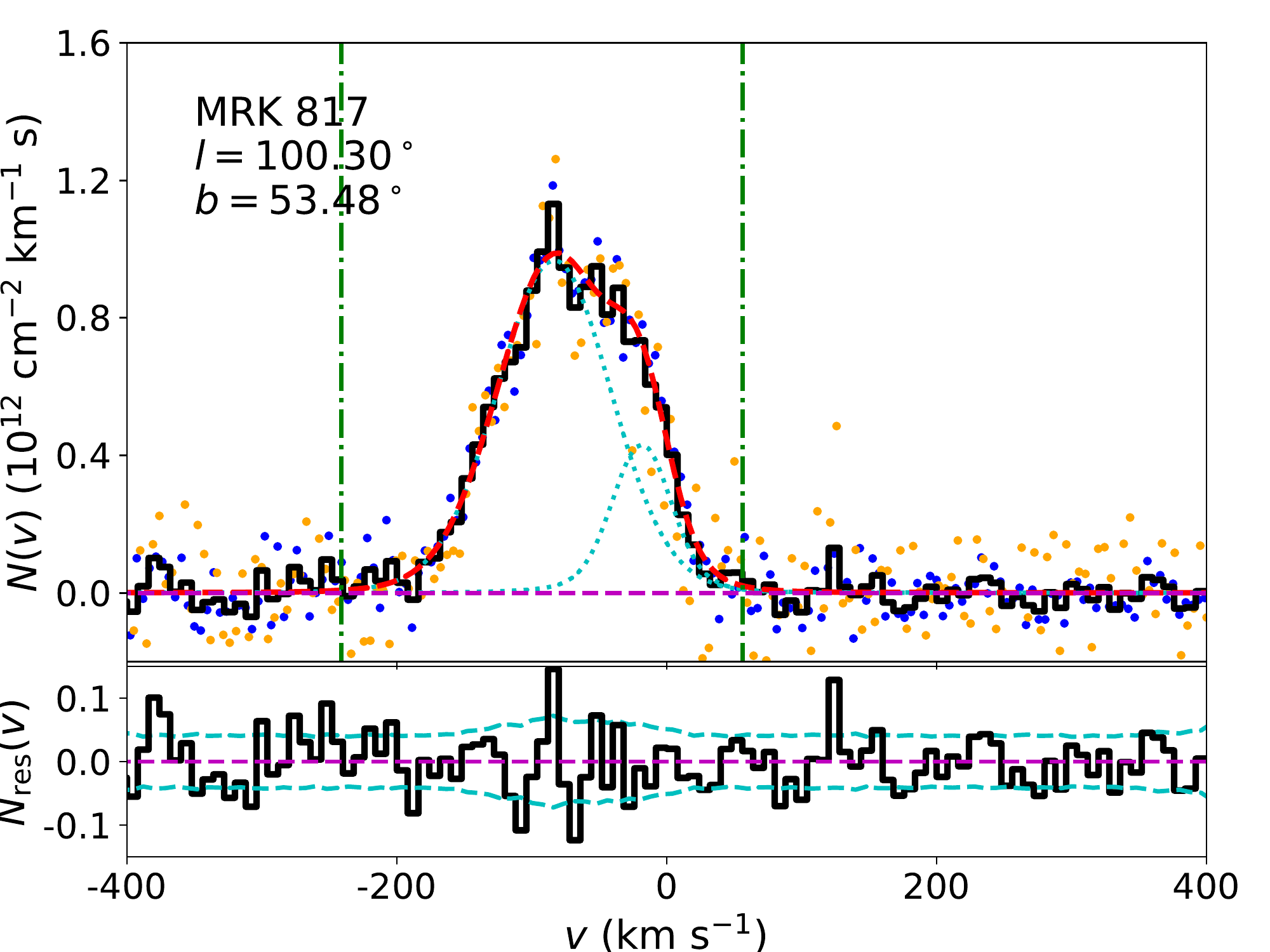}
\end{center}
\caption{Two example sight lines of PHL 2525 (left panel) and MRK 817 (right panel) showing combinations of the \ion{C}{4} doublet to obtain the column density line shapes (black and gray lines). In the upper panels, the blue and orange dots are the apparent column density of the strong ($1548.2 \rm\AA $) and the weak ($1550.8 \rm\AA $) lines, respectively. The cyan dotted lines show the individual Voigt components, while the red dashed line is the summation of all Voigt components. The lower panels show the Voigt fitting residuals (the black line), while the cyan dashed lines indicate the $1\sigma$ limits of uncertainty.}
\label{line_shape_ex}
\end{figure*}

There are three steps in this hybrid method.
We begin by fitting the doublet lines with the Voigt profile simultaneously, to extract the first order column density $N_{\rm Voigt}$.
The second step is to apply the AODM method to the residuals of the Voigt profile fitting to extract the residual column density $N_{\rm res}$ for both strong and weak lines.
Thirdly, we combine doublets for the final column density $N_{\rm tot} = N_{\rm Voigt}+N_{\rm res}$.
We reduce the matched \ion{Si}{4} sample with the same method to make this self-consistent, and find that the extracted column density line shapes are consistent with \citetalias{Qu:2020aa}.
Two examples of the line shape extraction are shown in Fig. \ref{line_shape_ex}.

We examine all of the Voigt fitting results manually.
The original spectrum from the HSLA is rebinned in 2 original pixels to increase $S/N$ per bin.
The continuum is fitted within $\pm 1000 \kms$ around the target doublets for both \ion{C}{4} and \ion{Si}{4}. 
The Voigt model is optimized by minimizing the $\chi^2$ through the Levenberg-Marquardt optimization.
Voigt components are added at the $2 \sigma$ level determined by the change of total $\chi^2$.
We note that the decomposition only affects the extraction of the central region (possibly saturated region) of the final column density line shape, so we test the reliability of the Voigt profile decomposition (discussed below).
Based on the Voigt model, we exclude HVCs associated with the Magellanic system \citep[MS; ][]{Fox:2014aa}, which cannot be modeled by the kinematical model \citepalias{Qu:2020aa}.
We obtain the column density line shape of the Voigt model $N_{\rm Voigt}(v)$.

Then, the AODM extraction is applied to the residual of the Voigt profile fitting to account for the remained variations:
\begin{equation}
N_{\rm res}(v) = -\frac{m_{\rm e}c}{\pi e^2 f \lambda} \ln \frac{I_{\rm obs}(v)}{I_{0}(v) I_{\rm m}(v)},
\end{equation}
where $N_{\rm res}(v)$ is the residual column density, $I_{0}(v)$, $I_{\rm m}(v)$, and $I_{\rm obs}(v)$ are the continuum flux, the Voigt model flux, and the observed flux.
We obtain the final line shape by adding the Voigt line profile with the AODM residual for both lines in the doublet (i.e., $N_{\rm tot}(v) = N_{\rm Voigt}(v) + N_{\rm res}(v)$).
The addition of the AODM residual not only introduces noises to the fitted Voigt profile, it also introduces the secondary variation, which is buried by noise.
This secondary variation could be crucial to capture the extended wings around the major absorption systems, which determines the density distribution at large radii in the kinematical model.
Using the MRK 817 in Fig. \ref{line_shape_ex} as an example, the absorption system is decomposed into two components based on the 2$\sigma$ threshold.
However, the major component at $-90 \kms$ may be decomposed further into another two narrower components, which is about 1 $\sigma$ better than the model shown.
The three-component Voigt profile model has a different blue wing (narrower) from the two-component decomposition.
This kind of difference will be accounted for by the AODM residuals, so the determination of the absorption wings in the hybrid method is insensitive to the decomposition of Voigt components.
The associated uncertainty of the column density has the same format as the AODM but normalizing the error flux with both the continuum and the Voigt model fluxes.
Therefore, the derived uncertainty is larger than the standard AODM, which accounts for the uncertainty introduced by the Voigt profile fitting.

The final column density profile is coadded in the new $8 \kms$ bins for \ion{C}{4} and $7.5 \kms$ for \ion{Si}{4}, which is about $3-4$ times the width of original pixels.
The strong and weak lines are weighted by the uncertainty for both \ion{C}{4} and \ion{Si}{4} doublets.
For the kinematical model, we only use the $-300$ to 300 $\kms$ region, after excluding the absorption features associated with the MS.

The robustness of this method depends on the reliability of the Voigt profile extraction, so we test the Voigt profile fitting program with saturation.
The \ion{C}{4} density distribution of individual components shows that most components ($95\%$) are less than $\log N = 14.3$, and the maximum column density is $\log N = 14.69$.
Then, we simulate the most limiting case by generating 100 random \ion{C}{4} doublets, where the continuum $S/N = 5$, and the column density and the velocity broadening are $\log N = 14.7$ and $b = 30\kms$.
Among these 100 random instances, the extracted column density deviations are mostly within $\sigma \approx 0.1$ dex.
However, there are about 10 instances that have column densities $\log N>15$, which are affected by saturation and the low $S/N$.
In practice, we notice that the highest column density components ($\log N > 14.6$) are all in relatively high $S/N > 10-15$ spectra.
In this case, the scatter of the extracted column density is $0.06$ dex and none of these extractions are affected by saturation.
For the majority of components ($S/N \approx 10$ and $\log N < 14.3$), the simulated column density scatter is $0.02$ dex, which is consistent with the measurement uncertainty.
Therefore, we suggest that the Voigt profile fitting in the hybrid method can define the first-order variation of the column density profile.

\begin{figure*}
\begin{center}
\includegraphics[width=0.49\textwidth]{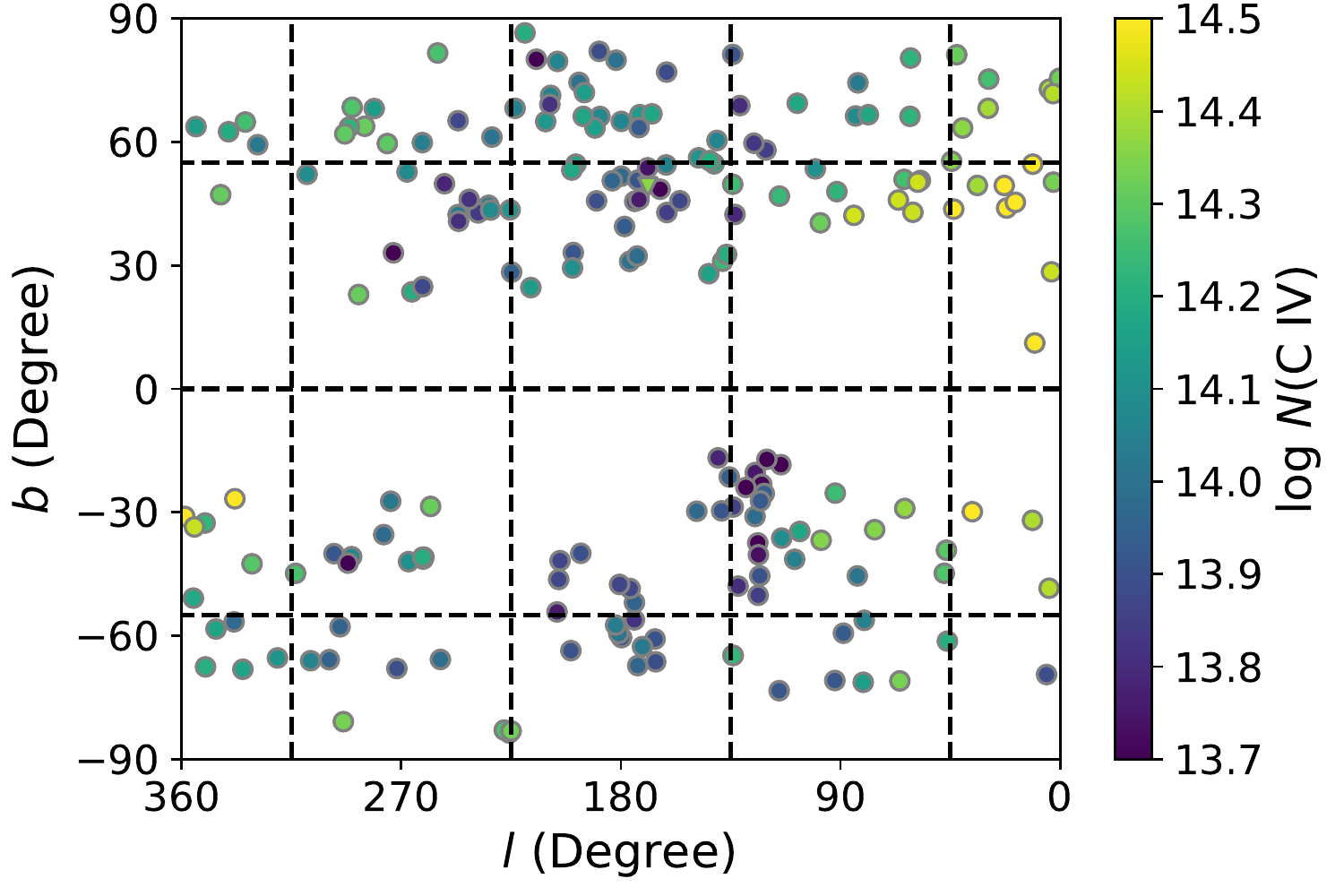}
\includegraphics[width=0.49\textwidth]{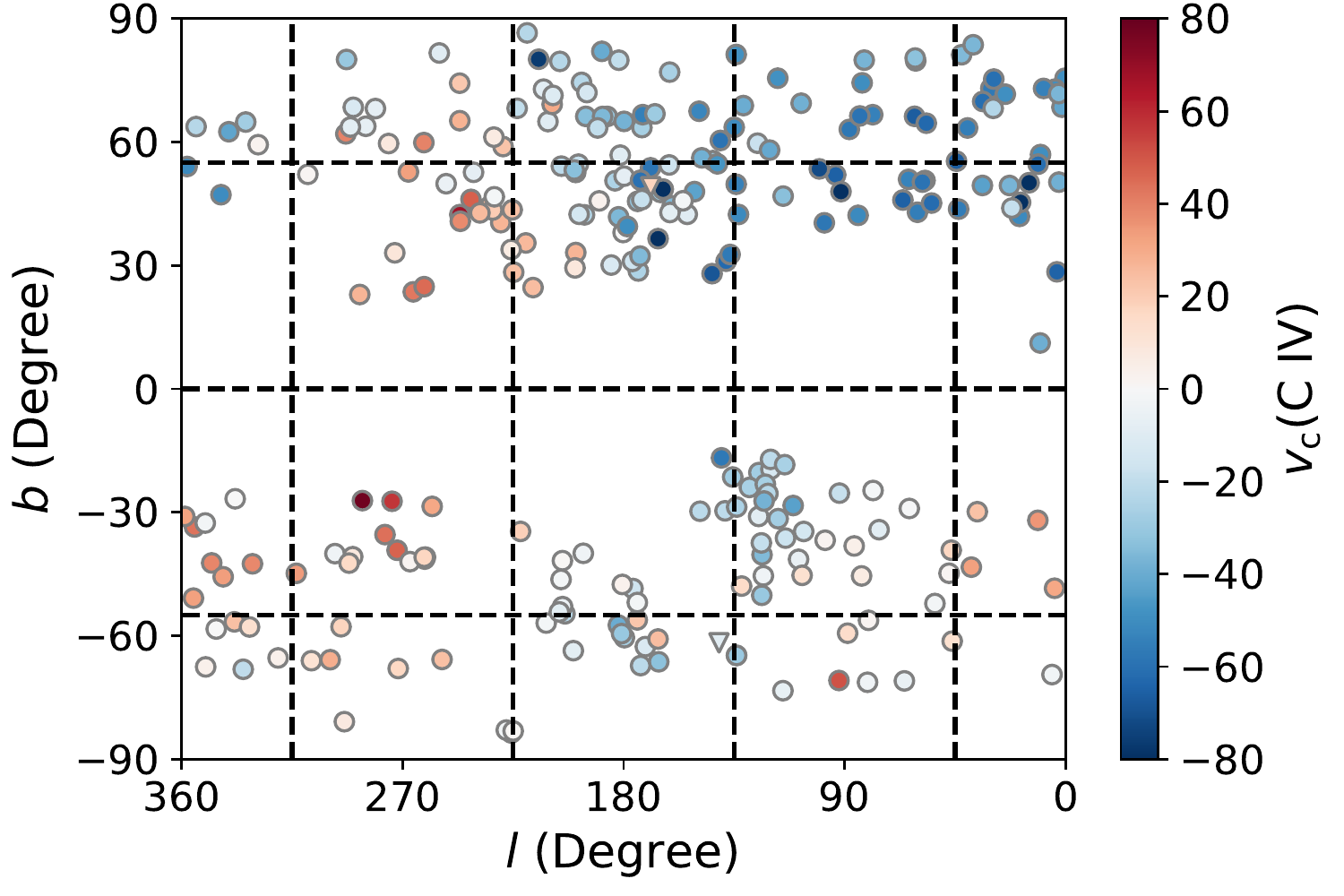}
\end{center}
\caption{The column density (left panel) and the line centroid (right panel) of the MW \ion{C}{4} line-shape sample. The column density is higher around the GC, which indicates the radial variation of the \ion{C}{4} density distribution. The line centroid map shows opposite velocities toward $90^\circ$ and $270^\circ$, which is induced by Galactic rotation.}
\label{logN_vc_c4}
\end{figure*}

We calculate the total column density and the line centroid for the MW absorption features (excluding the MS) from the extracted column density line shape.
The extracted velocity region depends on the width of Voigt profile fitting $2.8 b$, which is $\approx 4$ times the Gaussian width.
The results are summarized in Table \ref{sample} for both \ion{C}{4} and \ion{Si}{4}.
The total column density and line centroid of \ion{C}{4} are plotted in Fig. \ref{logN_vc_c4}, and similar plots for \ion{Si}{4} can be found in \citetalias{Qu:2020aa}.

\begin{table*}
\begin{center}
\caption{The Column Density Measurements of the Matched \ion{C}{4} and \ion{Si}{4} Sample}
\label{sample}
\begin{tabular}{cc|ccccc|ccccc|c}
\hline
\hline
 & & \multicolumn{5}{c|}{\ion{C}{4}} & \multicolumn{5}{c|}{\ion{Si}{4}} \\
    $l$ &     $b$ &  log$N$ & $\sigma_{\log N}$ & $v_{\rm c}$ & $\sigma_{v_{\rm c}}$ &  SN &  log$N$ & $\sigma_{\log N}$ & $v_{\rm c}$ & $\sigma_{v_{\rm c}}$ &  SN & Sight line \\
 (deg.) &  (deg.) &   (dex) & (dex) & km s$^{-1}$ & km s$^{-1}$ &     &   (dex) & (dex) & km s$^{-1}$ & km s$^{-1}$ &     &  \\
    (1) &     (2) &     (3) &   (4) &    (5) &   (6) & (7) &     (8) &   (9) &   (10) &  (11) & (12) & (13) \\
\hline
   0.24 &   75.52 &   14.32 &  0.04 &  -48.9 &   3.0 &   7 &   13.62 &  0.04 &  -46.0 &   3.8 &   7 & SDSSJ134246.89+184443.6 \\
   1.51 &   68.48 &   14.30 &  0.04 &  -41.4 &   2.5 &   7 &   $...$ & $...$ &  $...$ & $...$ & $...$ & VV2006-J140658.8+144238 \\
   2.75 &   50.26 &   14.33 &  0.01 &  -38.3 &   0.5 &  23 &   13.79 &  0.01 &  -38.5 &   1.2 &  24 & MRK1392 \\
   2.86 &   71.78 &   14.41 &  0.02 &  -36.7 &   2.2 &  10 &   13.74 &  0.01 &  -43.5 &   1.3 &  29 & SDSSJ135712.61+170444.1 \\
   3.53 &   28.46 &   14.43 &  0.02 &  -64.6 &   3.1 &  11 &   13.89 &  0.02 &  -49.9 &   4.8 &   9 & 1H1613-097 \\
   4.37 &   72.87 &   14.40 &  0.01 &  -48.9 &   1.8 &  22 &   13.68 &  0.01 &  -47.0 &   0.9 &  44 & PG1352+183 \\
   4.51 &  -48.46 &   14.41 &  0.02 &   31.0 &   1.4 &  10 &   13.65 &  0.02 &   30.8 &   2.2 &  21 & RXS-J21388-3828 \\
\hline
\hline
\end{tabular}
\end{center}
Columns: (1) Galactic longitude; (2) Galactic latitude; (3) and (4) are the \ion{C}{4} column density and its uncertainty; (5) and (6) are the Weighted \ion{C}{4} line centroid and its uncertainty; (7) the $S/N$ ratio at the \ion{C}{4} doublet; (8) and (9) are the \ion{Si}{4} column density and its uncertainty; (10) and (11) are the Weighted \ion{Si}{4} line centroid and its uncertainty; (12) the $S/N$ ratio at the \ion{Si}{4} doublet; (13) the sight line name.
The entire version of this table is in the online version.
\end{table*}

\section{The Kinematical Model of \ion{C}{4}-bearing Gas}
\subsection{Overview}

With the line shape sample, we adopt a kinematical model to decompose the density profile and the bulk velocity for the \ion{C}{4}-bearing gas.
The basis of the method is the kinematical distance -- the projected radial velocity of an absorption system can be converted to a distance with a bulk velocity field (e.g., rotation).
The physical model and Bayesian fitting model are described in \citetalias{Qu:2020aa} (Table 2), which summarizes all parameters in the kinematical model.
Here, we briefly introduce the kinematical model.

The warm gas density distribution is assumed to be the summation of an ``exponential" disk and a $\beta$-model CGM ($n = n^{\rm disk}+n^{\rm CGM}$).
The ``exponential" disk has a dependence in both $z$ (zenith direction) and $r_{\rm xy}$ (midplane radial direction), given by  $n^{\rm disk}(r_{\rm xy}, z) = n_0^{\rm disk} \exp[-(r_{\rm xy}/r_0)^{\alpha_{\rm r}}-(z/z_0)^{\alpha_{\rm z}}]$, where $\alpha_{\rm xy}$ and $\alpha_{\rm r}$ are the slope in the exponential function.
The CGM component is assumed to be an anisotropic $\beta$-model of the form $n^{\rm CGM}(r, \theta) = n_0^{\rm CGM}(\theta) \times (1+(\frac{r}{r_{\rm c}})^2)^{-3\beta/2}$, where $\theta$ is the elevation angle related to the GC (Fig. 4 in \citetalias{Qu:2020aa} for an illustration).
At the small scale, the warm gas is assumed to be separate clouds, because previous UV absorption line studies show that the warm gas cloud is clumpy.
The traditional way to model this clumpiness is by adopting a patchiness parameter (additional scatter of the column density per sight line; \citealt{Savage:2009aa} and reference therein).
In \citetalias{Qu:2020aa}, the cloud nature suggests this patchiness is induced by the Poisson uncertainty of the number of clouds along the line of sight.
Then we constrain the characteristic column density per cloud ($\log N_{\rm sg}$) in the kinematical model.
For individual clouds, another empirical parameter is the warm gas broadening velocity ($v_{\rm abs}$), which is the combination of the intrinsic thermal broadening,  the turbulent broadening, and the instrumental broadening.

There are also two contributors to the bulk velocity field: the Galactic rotation and the radial velocity (i.e., infall and outflow).
The Galactic rotation is assumed to be a cylindrical rotation with a velocity of $v_{\rm rot} (r, \theta) = v_{\rm rot}(r_{\rm XY}) \cos \theta$, where $r_{\rm XY}$ is the radius in the midplane (the XY plane).
For the radial velocity, we assume a constant mass ejection or accretion rate, so there is a radial dependence of the radial velocity.
The radial velocity is only calculated beyond a disk boundary (i.e., the disk component is rotation-only), which is parameterized in the disk shape with a constant of $R = -(r_{\rm xy}/r_0)^{\alpha_{\rm r}}-(z/z_0)^{\alpha_{\rm z}}$.
In addition to the bulk velocity, the warm gas shows residual inflows and outflows, which is modeled by a parameter $v_{\rm rand}$ for these random residuals.

The Bayesian implementation of the kinematical model is mostly the same as \citetalias{Qu:2020aa}.
The only difference is for the calculation of the velocity difference between the model and the observed line shape.
In \citetalias{Qu:2020aa}, the velocity difference is calculated from the cross-correlation spectrum.
In this work, we update this calculation by the direct difference between the line centroid of the model and observation. 
By using this implementation, we could also include the uncertainty of the measured line centroid in the likelihood calculation:
\begin{equation}
\ln p_{\rm rand} = - \ln v'_{\rm rand} - \frac{1}{2} (\frac{v_{\rm shift}}{v'_{\rm rand}})^2,
\end{equation}
where $v'_{\rm rand}= (v_{\rm rand}^2 + \sigma_{v}^2)^{1/2}$, and $\sigma_{v}$ is the velocity measurement uncertainty.

\begin{figure*}
\begin{center}
\includegraphics[width=0.99\textwidth]{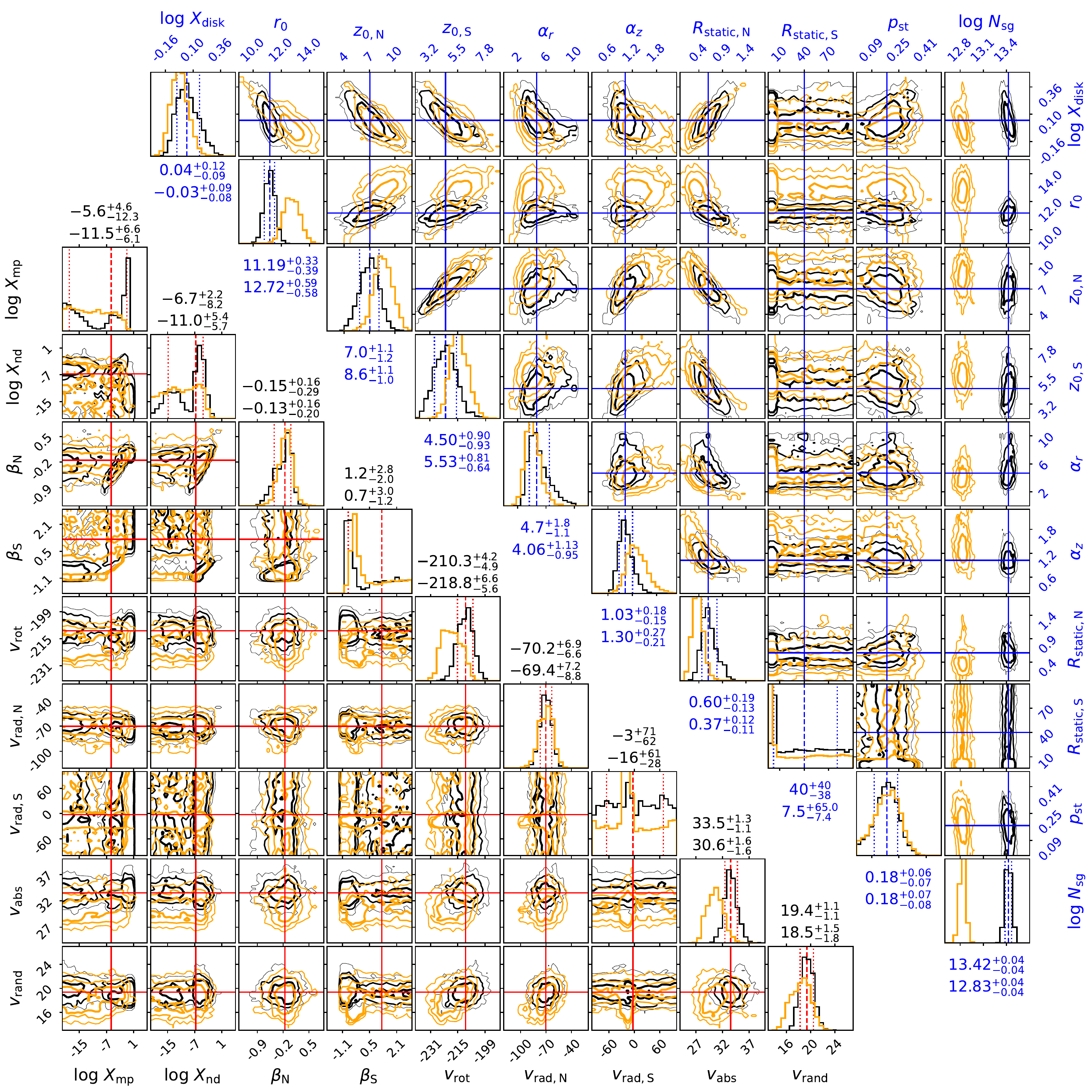}
\end{center}
\caption{The posterior distribution of the parameters in the \ion{C}{4} (black contours) and \ion{Si}{4} (orange contours) kinematical models.
The definition of parameters can be found in the text and Table 2 in \citetalias{Qu:2020aa}.
All parameters are consistent between \ion{C}{4} and \ion{Si}{4} within $2 \sigma$, except for the column density per cloud ($\log N_{\rm sg}$).
The $\log N_{\rm sg}$ value of \ion{Si}{4} is 0.6 dex lower than the \ion{C}{4} value, which is about the median column density ratio between \ion{Si}{4} and \ion{C}{4} (Section 4.1).
We note that the CGM component of \ion{C}{4} shows two possible solutions (i.e., $\log X_{\rm mp}$ and $\log X_{\rm nd}$), and details of these two solutions are plotted in Fig. \ref{params_2sol}.
}
\label{params_corner}
\end{figure*}

The kinematical model is applied to both the \ion{Si}{4} and the \ion{C}{4} samples.
The best solutions are extracted by the Markov chain Monte Carlo simulation using the {\it emcee} package \citep{Foreman-Mackey:2013aa}.
The fitted parameters of \ion{Si}{4} are consistent with those in  \citetalias{Qu:2020aa} within $\approx 2 \sigma$ (Fig. \ref{params_corner}).
The three parameters with the largest differences are $\alpha_{r}$, $\alpha_{z}$ and $\beta_{\rm N}$, which are affected by the modification of the velocity shift (discussed in Section 3.2).

The \ion{C}{4} modeling is consistent with \ion{Si}{4}, where most parameters are within $2 \sigma$.
The most significant difference is the column density per cloud ($\log N_{\rm sg}$), which is different by 0.6 dex.
This difference is consistent with the difference between \ion{C}{4} and \ion{Si}{4} total column densities, which corresponds to similar number densities of clouds for both \ion{C}{4} and \ion{Si}{4}.
This difference is mainly due to the higher abundance of carbon, which makes \ion{C}{4} more sensitive to the low-density gas at large radii.
For \ion{Si}{4}, only upper limits are obtained for the CGM components $\log X_{\rm mp}$ and $\log X_{\rm nd}$, which are the number density of separated clouds along the midplane direction (mp) and the normal line direction (nd), respectively.
The model of \ion{C}{4} shows two possible solutions determined by relative values between $\log X_{\rm mp}$ and $\log X_{\rm nd}$, which is not observed in the \ion{Si}{4} modeling.
One solution suggests a disk that is more extended along the $z$-direction with $\log X_{\rm mp} \lesssim -8$, which is smaller than $\log X_{\rm nd} \approx -5.3$.
Another solution is more extended along the radial direction with $\log X_{\rm mp} \approx -1.2$, which is  larger than $\log X_{\rm nd} \lesssim -8$.
These two solutions show similar probabilities in the posterior distribution, which is limited by the current \ion{C}{4} and \ion{Si}{4} samples with few low Galactic latitude sight lines (e.g., $\lesssim 25^\circ$).
Details of these two solutions are discussed in Section 3.2.

Although the \ion{C}{4} model is consistent with the \ion{Si}{4} model within $2 \sigma$ (Fig. \ref{params_corner}), it does not mean that these two ions share the exactly same global variation.
There is still a large-scale difference, which shows up when we consider the difference of the column density between \ion{C}{4} and \ion{Si}{4} (Section 4.1). 

\subsection{Density Distribution and Kinematics}
The kinematical modeling of \ion{C}{4} absorption features suggests two possible solutions, so we consider both solutions for their density distributions and kinematics.
Phenomenologically, one solution is more extended along the zenith direction ($z$) of the disk, which is suggested in \citetalias{Qu:2020aa}.
Another solution is more extended along the radial direction of the disk, which may explain the lack of the warm CGM in the MW, compared to the COS-Halo $L^*$ galaxy survey at $z\approx 0.2$ \citep{Bish:2021aa}.
More low Galactic latitude sight lines are needed to distinguish between these two solutions.
Here we consider the density distributions and kinematics of these two solutions for the current sample.

\begin{figure}
\begin{center}
\includegraphics[width=0.49\textwidth]{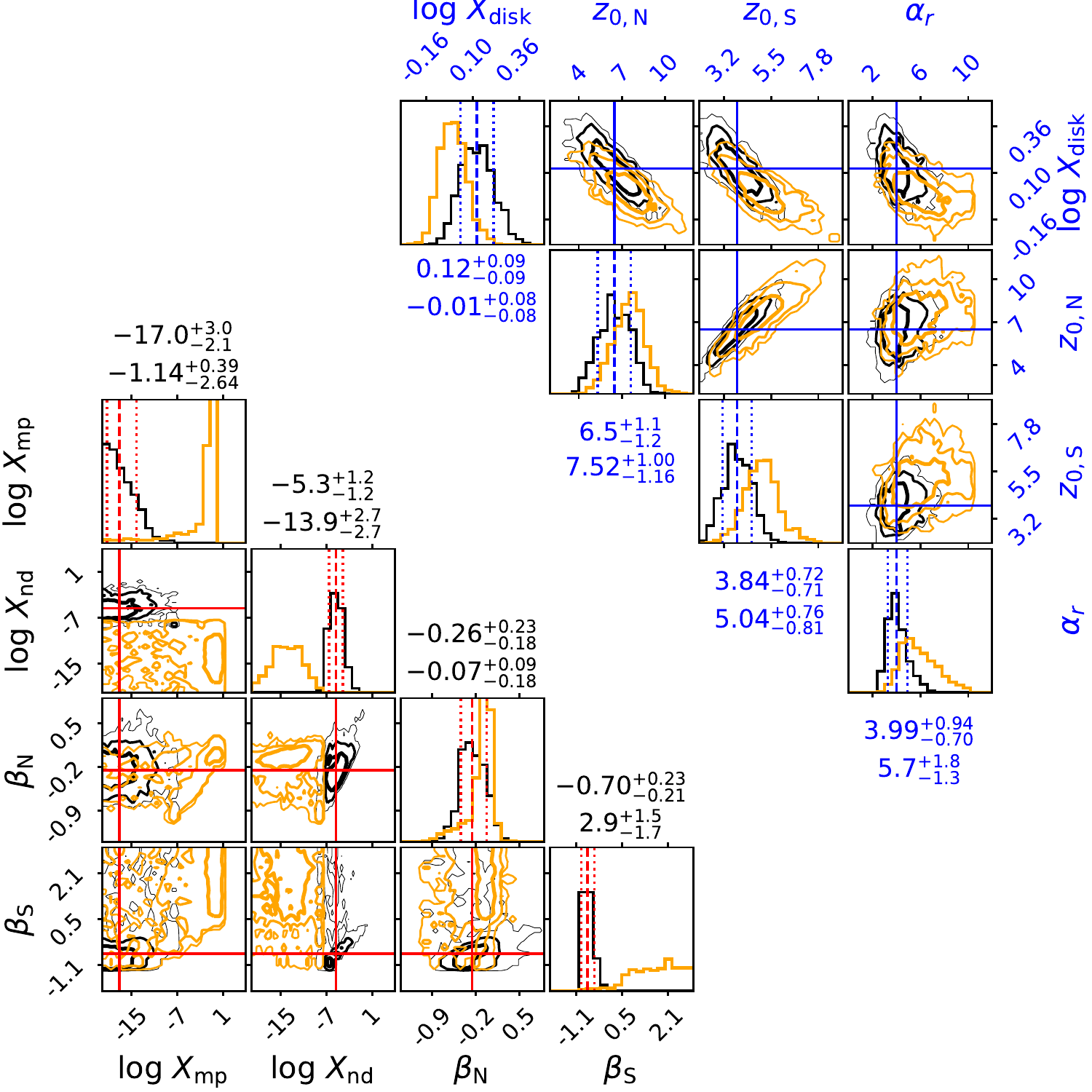}
\end{center}
\caption{The parameters showing $> 0.5\sigma$ difference between two possible solutions of the \ion{C}{4} density distribution. We adopt a boundary of $\log X_{\rm nd} = -8.2$ for these two solutions: Model I of $\log X_{\rm nd} \geq -8.2$ and Model II of $\log X_{\rm nd} < -8.2$.}
\label{params_2sol}
\end{figure}

As shown in Fig. \ref{params_corner}, these two solutions are well separated in the joint posterior distribution of  $\log X_{\rm nd}$ and $\log X_{\rm mp}$.
Here, we adopt a boundary of $\log X_{\rm nd} = -8.2$ for these two solutions: Model I of $\log X_{\rm nd} \geq -8.2$ and Model II of $\log X_{\rm nd} < -8.2$.
In Fig. \ref{params_2sol}, we show all parameters with differences $> 0.5 \sigma$.
These differences are all associated with the density distribution rather than the kinematics.

\begin{figure*}
\begin{center}
\includegraphics[width=0.48\textwidth]{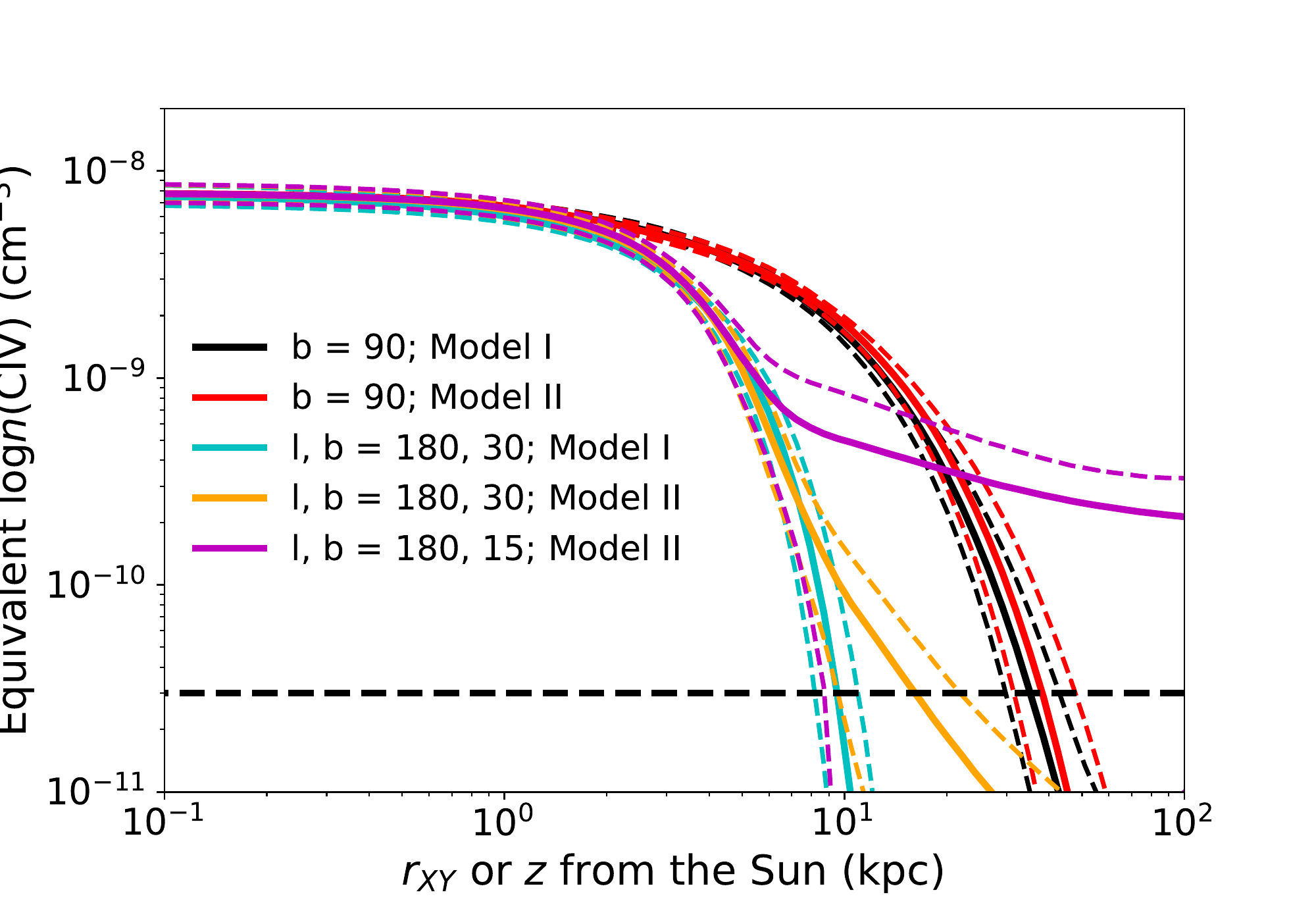}
\includegraphics[width=0.48\textwidth]{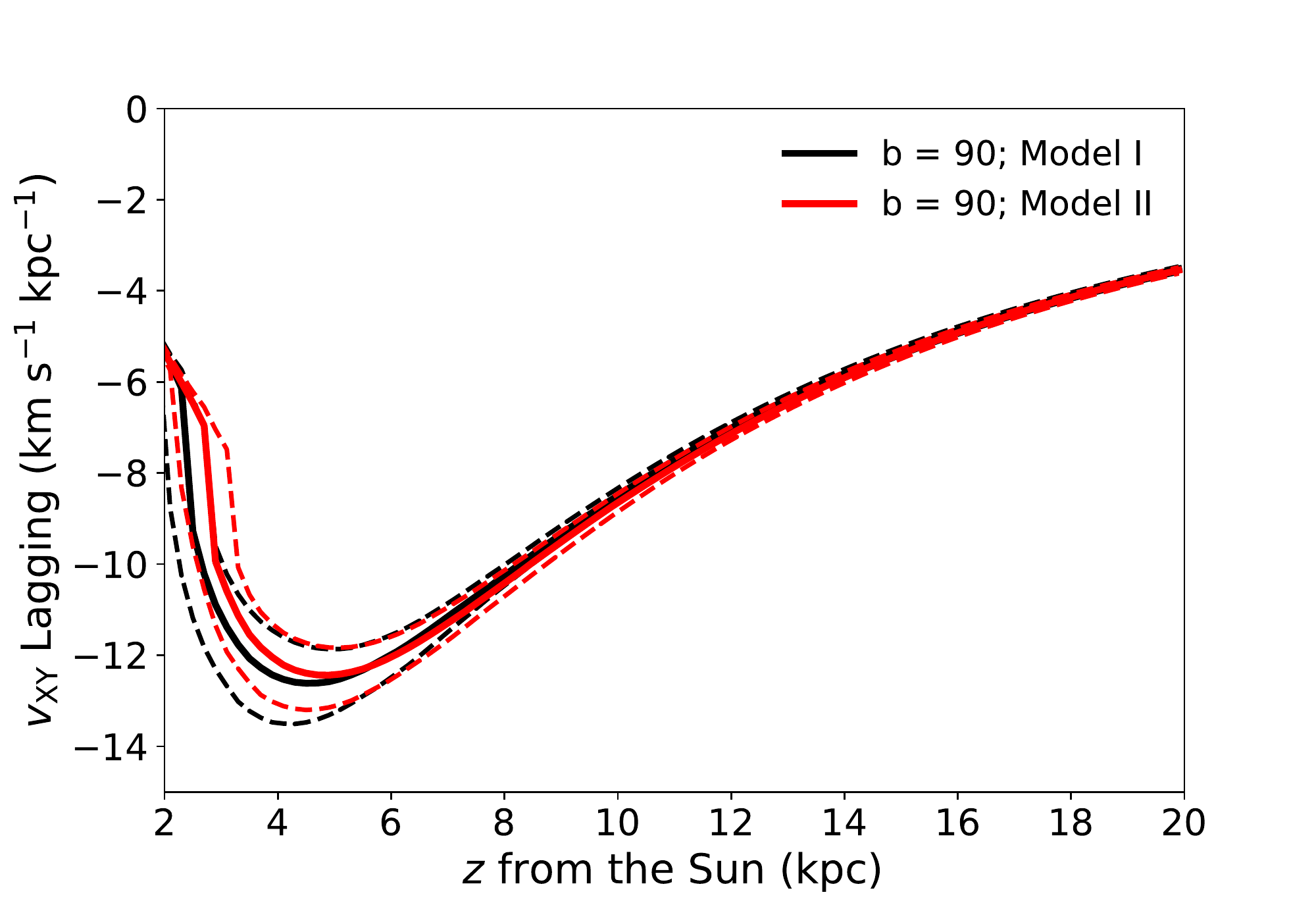}
\end{center}
\caption{The equivalent density distribution (left panel) and the rotation lagging close to the disk (right panel) of the MW \ion{C}{4}.
The density distribution of the Model I and II are consistent for the current sample ($b \gtrsim 30^\circ$).
However, at low Galactic latitude ($15^\circ$ in the plot), Model II will be much more extended, but it cannot be distinguished by the current sample.
The vertical dashed line represents the detection limit for the current sample at high latitudes.
The right panel shows that the density distribution does not affect the determination of the lag relative to Galactic rotation.}
\label{logn_vlag}
\end{figure*}

As argued in \citetalias{Qu:2020aa}, it is the summation of the disk and CGM components to approximate the real density distribution.
We examine whether these two models present consistent total density distributions in Fig. \ref{logn_vlag}.
The median and $1 \sigma$ uncertainties are extracted from the MCMC chain.
In this plot, the detection limits are estimated from the continuum $S/N$ and the number of sight lines, which is about $3\times 10^{-11} \cc$ for \ion{C}{4}.
It is clear that the total density profiles in the two solutions are consistent with each other within $1 \sigma$ around the Solar system for the $z$-direction of the disk, and the low latitude sight lines ($|b|= 30^\circ$).
As a comparison, we plot the $|b|=15^\circ$ sight line for Model II, which shows an extended disk along the radial direction.
In Model II, the high density at large radii leads to a significant contribution to the total column density at high velocities of $|v_{\rm LSR}| \approx 100 \kms$ at $|b|=15^\circ$, which can be confirmed or ruled out by observation at low latitudes.
However, the archival {\it HST}/COS \ion{C}{4} sample only has 5/265 sight lines with $b < 20^\circ$.

In the \ion{C}{4} model, there is an issue for the $\beta$ values, the slope of the power-law density distribution at large radii, which is expected to be positive (i.e., density decreases at large radii).
However, as shown in Fig. \ref{params_corner}, the $\beta$ value has a median of $-0.15$ in the northern sky.
For the southern sky, the two solutions show significantly different $\beta$ values of $<0.0$ and $>0.5$.
This is because in the kinematical model the density distribution is not the only factor affecting the $\beta$ value, and because we assume a constant accretion or outflow mass rate for the radial velocity.
This assumption implies that the $\beta$ value is affected by both the density distribution and the bulk velocity.
We examine that the determination of the $\beta$ value is dominated by the kinematics close to the disk rather than the density distribution at large radii.
In Fig. \ref{logn_vlag}, we show that the two solutions have similar velocities in the XY plane ($v_{\rm XY}$), which is a combination of the rotation velocity and the projection of the radial velocity.
The extracted velocity lag relative to the disk midplane is about $-10 \kms~kpc^{-1}$ within 10 kpc.

We estimate the mass of the \ion{C}{4}-bearing gas by integrating the density distribution and correcting for the ion abundance.
Following the same steps as \citetalias{Qu:2020aa}, the disk component has a total mass of $\log M({\rm CIV}) = 7.96_{-0.03}^{+0.04} - \log( f/0.1) - \log(Z/0.5 Z_\odot)$, where $0.1$ and $0.5Z_\odot$ are the assumed ionization fraction and the metallicity.
The two solutions have consistent masses of $\log M({\rm CIV}) = 7.96\pm 0.03$ and $\log M({\rm CIV}) = 7.96_{-0.03}^{+0.05}$.
For the CGM components, although the median $\beta$ is negative, the mass is limited by the low cloud path length density (i.e., $\log X_{\rm mp}$ and $\log X_{\rm nd}$).
The $3 \sigma$ upper limits for the mass are $\log M({\rm CIV}) < 9.3$ and $9.2$ for Model I and II, respectively.
Both of these estimations are consistent with the \ion{Si}{4} gas with the disk mass of $\log M({\rm SiIV}) = 8.09_{-0.04}^{+0.05} - \log( f/0.2) - \log(Z/0.5 Z_\odot)$ and the CGM mass of $\log M({\rm SiIV}) < 9.1$.

For the bulk velocity, most of the related parameters are consistent between \ion{C}{4} and \ion{Si}{4} as shown in Fig. \ref{params_corner}.
Then, the extracted kinematics for \ion{C}{4} show similar global features as \ion{Si}{4} for two major components with $1 \sigma$.
First, the \ion{C}{4}-bearing gas is co-rotating with the stellar disk at a midplane velocity $210-220 \kms$.
At higher $z$, we also examine the velocity lag, which is a combination of both the rotation and the radial velocity components, since the radial velocity is projected onto the Galactic XY plane.
In Fig. \ref{logn_vlag}, we show the extracted lag velocity of the $z$-height, which is about $-10\kms~ kpc^{-1}$ close to the disk.
Second, the net inflow velocity in the northern hemisphere is about $-70\kms$ for both \ion{Si}{4} and \ion{C}{4}.
The total mass accretion rates are $-0.44_{-0.24}^{+0.11} ~M_\odot \rm ~ yr^{-1}$ and $-0.37_{-0.33}^{+0.10}~M_\odot \rm ~ yr^{-1}$ for Model I and II, respectively.
These accretion rates are all dominated by the northern hemisphere, which are $-0.42_{-0.14}^{+0.08}$ and $-0.38_{-0.08}^{+0.06}$ for Model I and II.
The southern hemisphere shows a $2\sigma$ upper limit of $1.3 M_\odot ~ \rm yr^{-1}$.

The similarity of both density and kinematics suggests that \ion{Si}{4} and \ion{C}{4} are cospatial at a large scale with roughly the same global features.
This is due to the close phases of these two ions.
However, there are also small-scale variations for these two ions, which are mainly discussed in Section 4.

\subsection{Model-Corrected Variations}
As argued in \citetalias{Qu:2020aa}, there are still small-scale features after subtracting the kinematical model from the observation.
In this work, we consider two means to characterize the small scale features -- the kinematical model correction residuals (this section) and the difference map between \ion{Si}{4} and \ion{C}{4} (Section 4).
As an improvement to \citetalias{Qu:2020aa}, we introduce a quantitative method to model the absolute difference and the auto-correlation coefficient.

We will study the difference maps in the following analysis, so we use the joint sample of both \ion{Si}{4} and \ion{C}{4} to have an apples-to-apples comparison with the kinematical model correction.
The joint sample has 203 sight lines with continuum $S/N > 5$, but there are still sight lines with large measurement uncertainties.
Then, we select sight lines with velocity uncertainties of $< 50 \kms$ (mostly $<20 \kms$) for both \ion{Si}{4} and \ion{C}{4} and column density uncertainties of $< 0.15$ dex to exclude the most uncertain measurements.
In the final sample, there are 198 sight lines for the joint analysis of \ion{Si}{4} and \ion{C}{4}.

\begin{figure*}
\begin{center}
\includegraphics[width=0.48\textwidth]{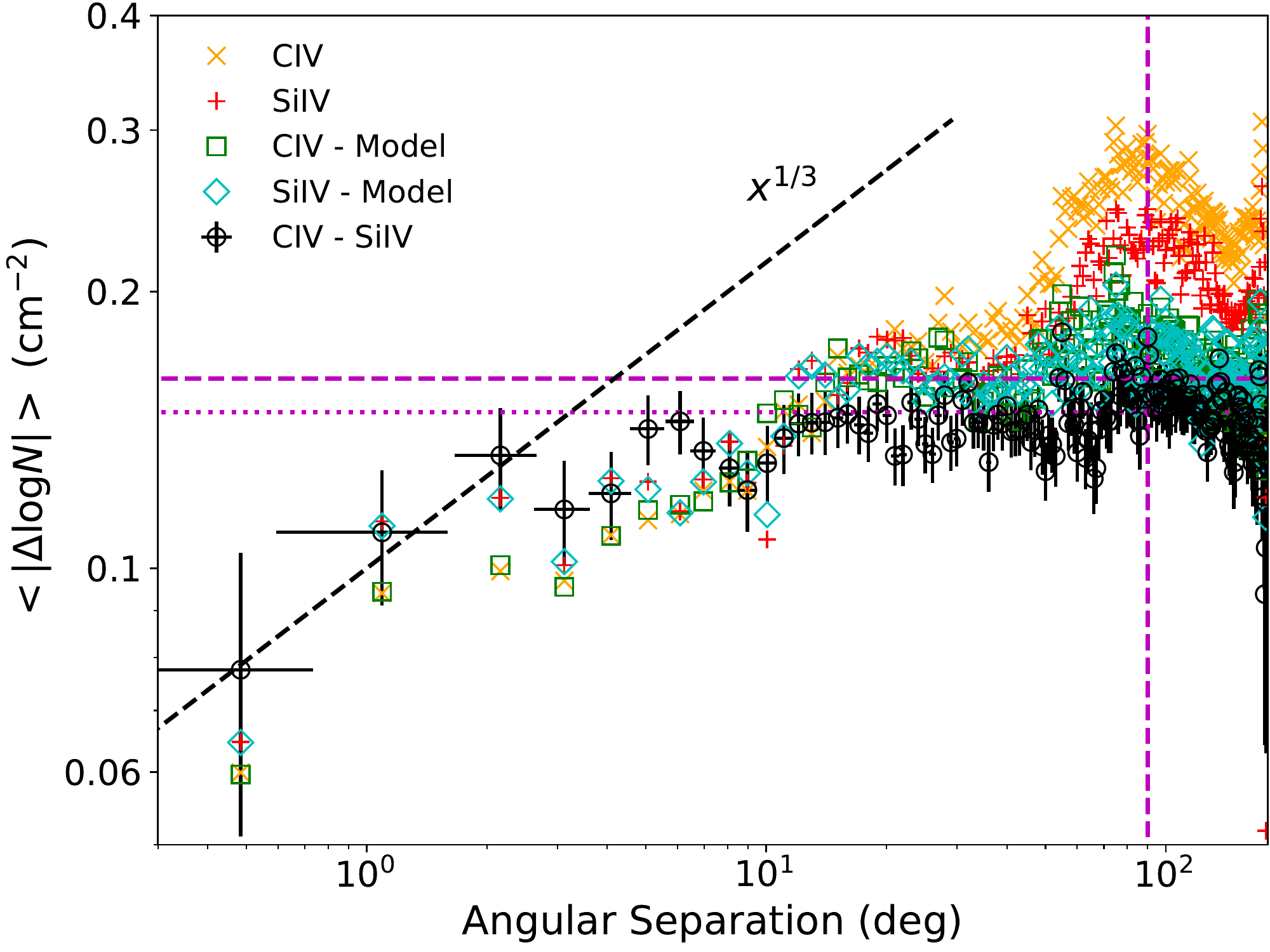}
\includegraphics[width=0.48\textwidth]{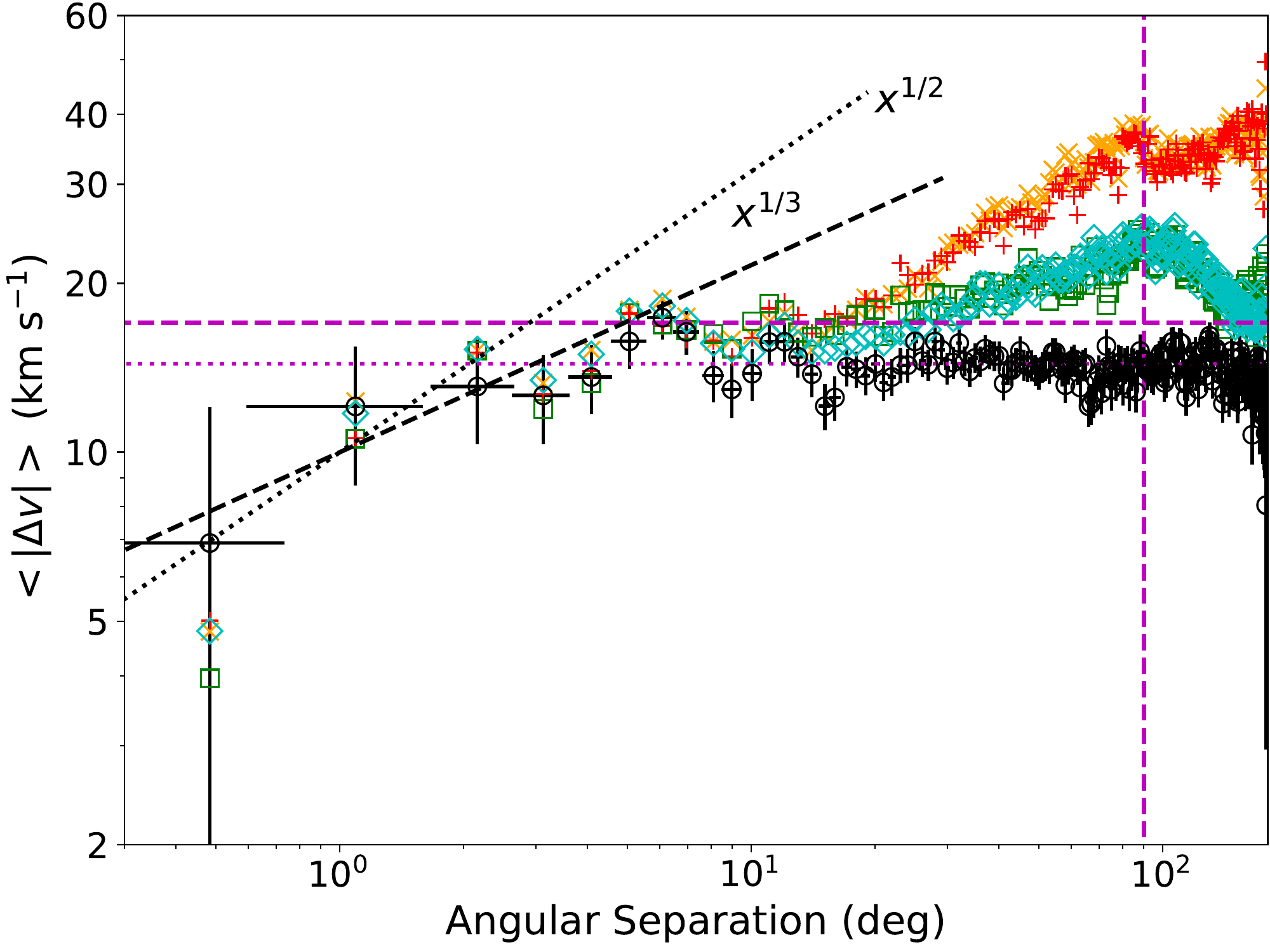}
\end{center}
\caption{The column density structure function (left panel) and the velocity structure functions (right panel).
In both panels, the original \ion{C}{4} and \ion{Si}{4} extractions are shown in orange crosses and red pluses, respectively, while the kinematical model-corrected residual are plotted in green squares and cyan diamonds.
The extractions for the difference map are shown in black circles with uncertainties, which are determined from the bootstrap method (similar for different extractions).
It is clear that the kinematical model captures the large-scale features for both the column density and the kinematics, so the large-scale variation is suppressed.
The difference map further suppresses feature at $\gtrsim 10^\circ$, especially for the velocity structure function.
This means that these two ions are not only cospatial globally (e.g., the rotating disk), they also share similar inflow and outflow features.
}
\label{logN_vc_sf}
\end{figure*}

\begin{figure*}
\begin{center}
\includegraphics[width=0.48\textwidth]{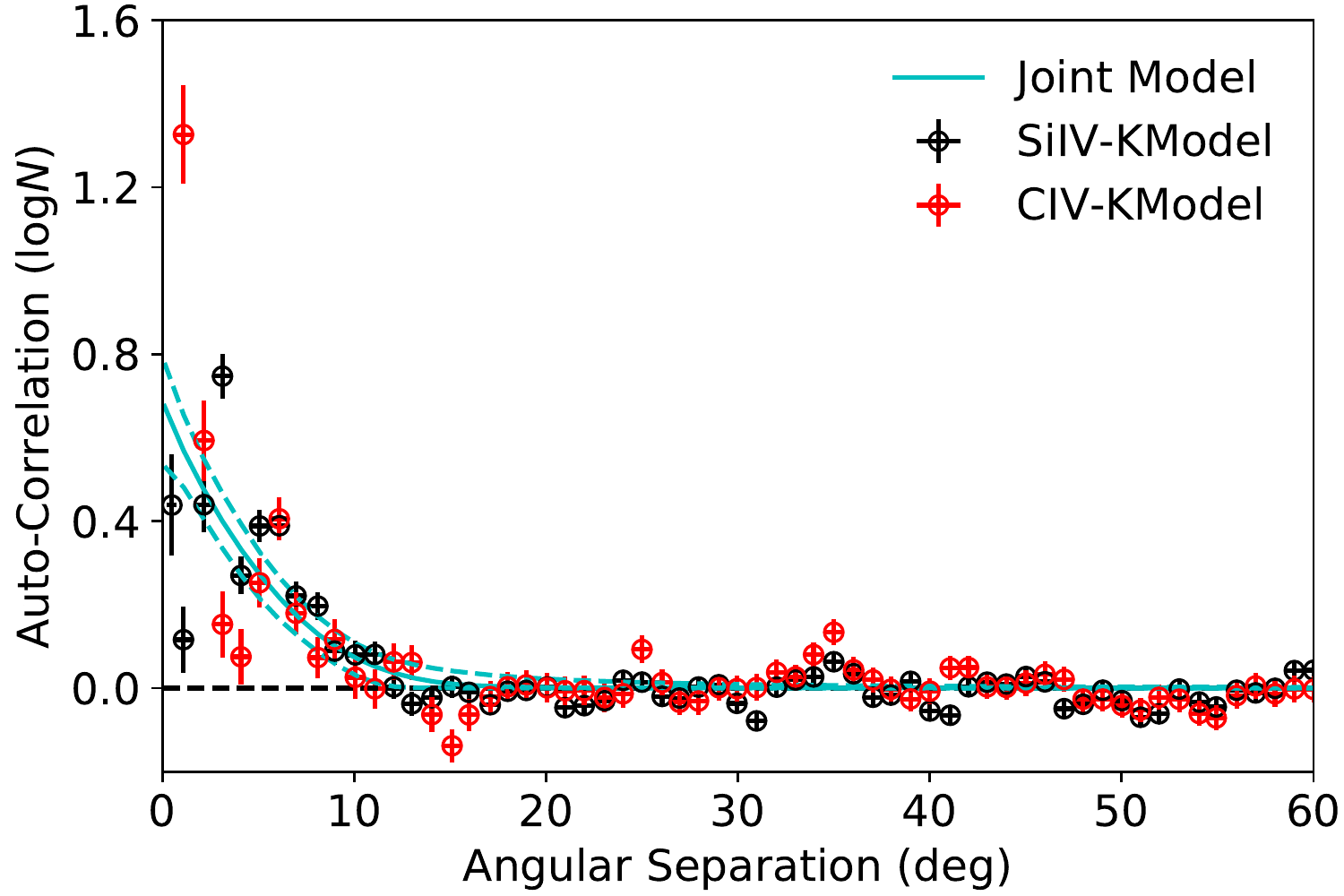}
\includegraphics[width=0.48\textwidth]{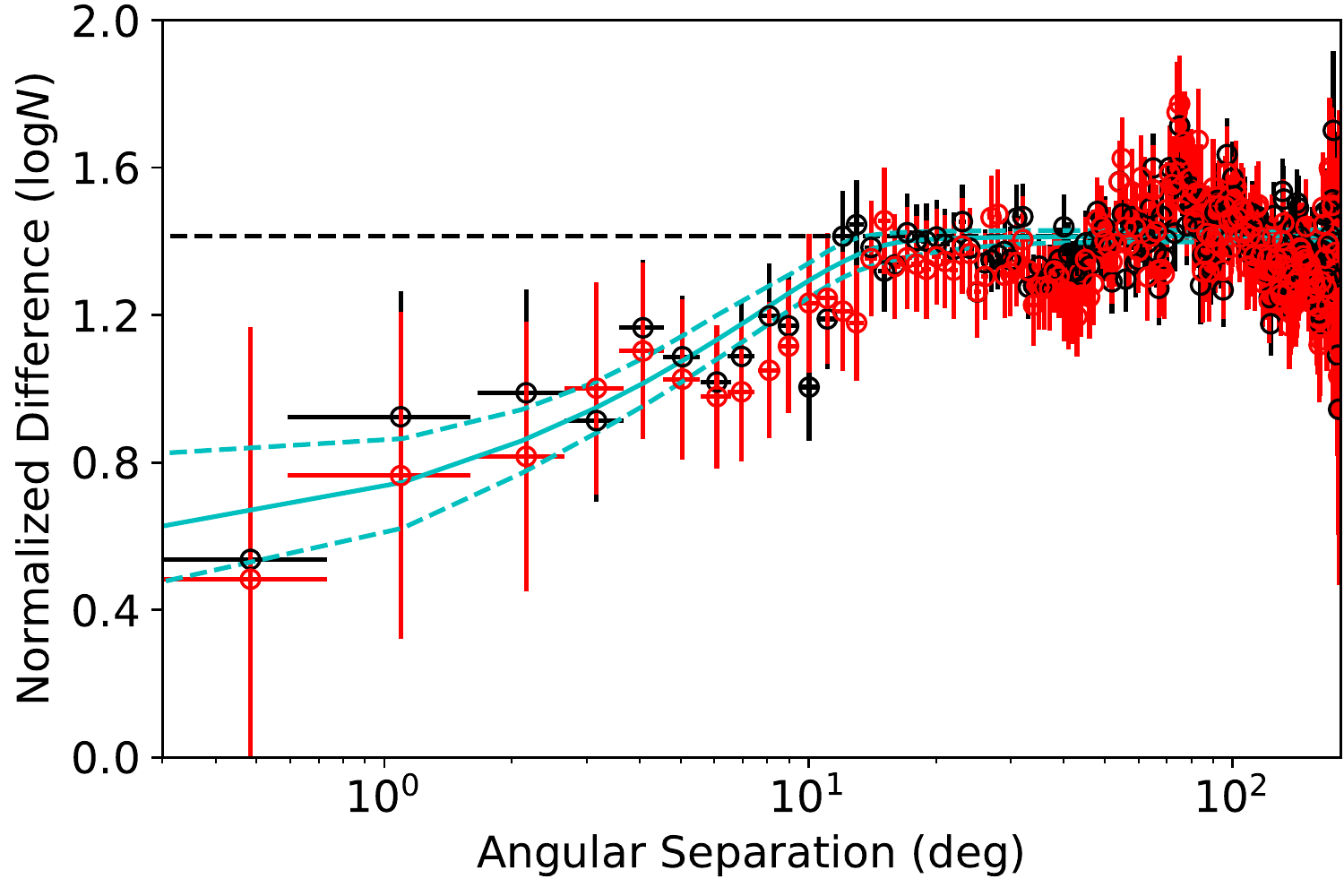}
\includegraphics[width=0.48\textwidth]{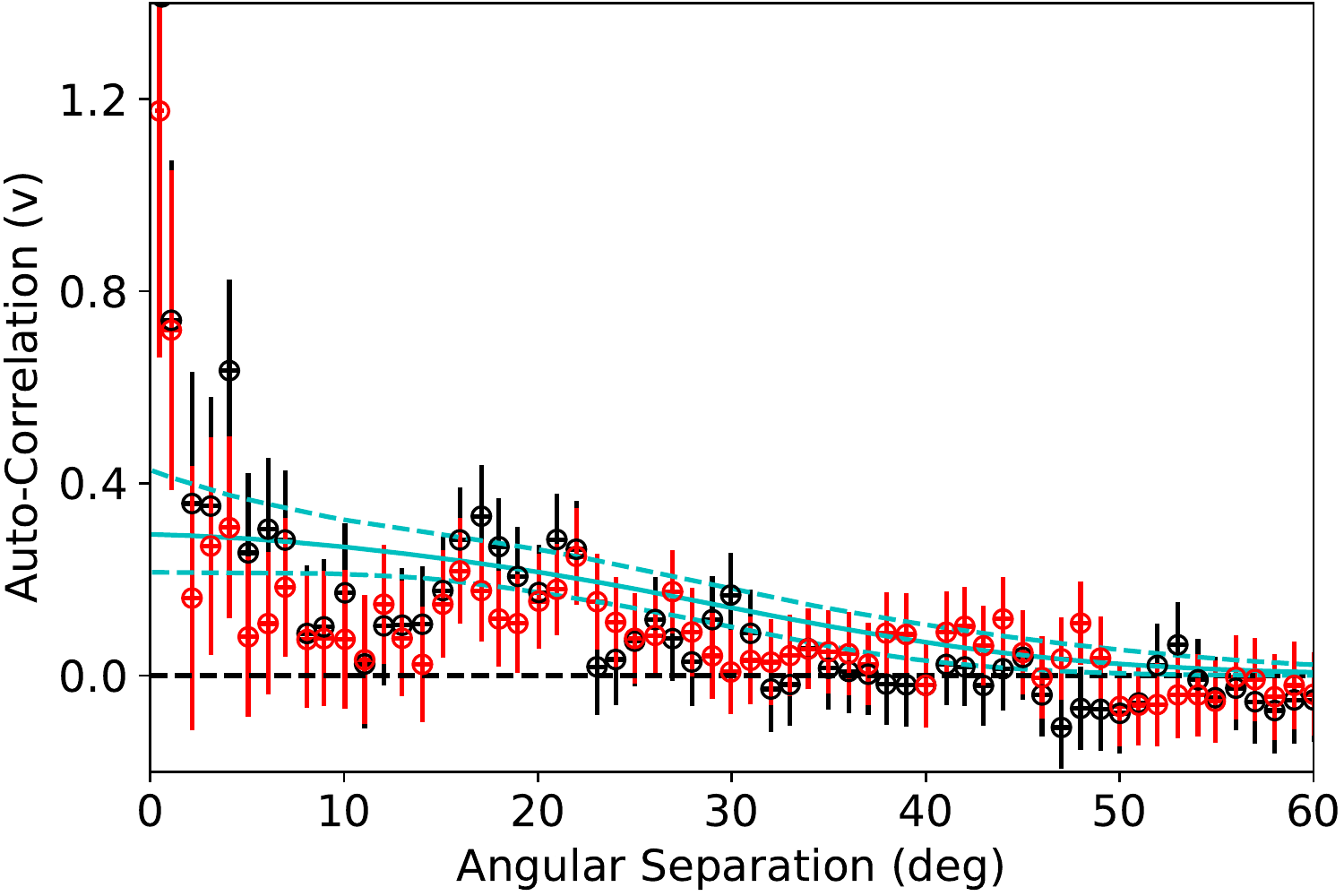}
\includegraphics[width=0.48\textwidth]{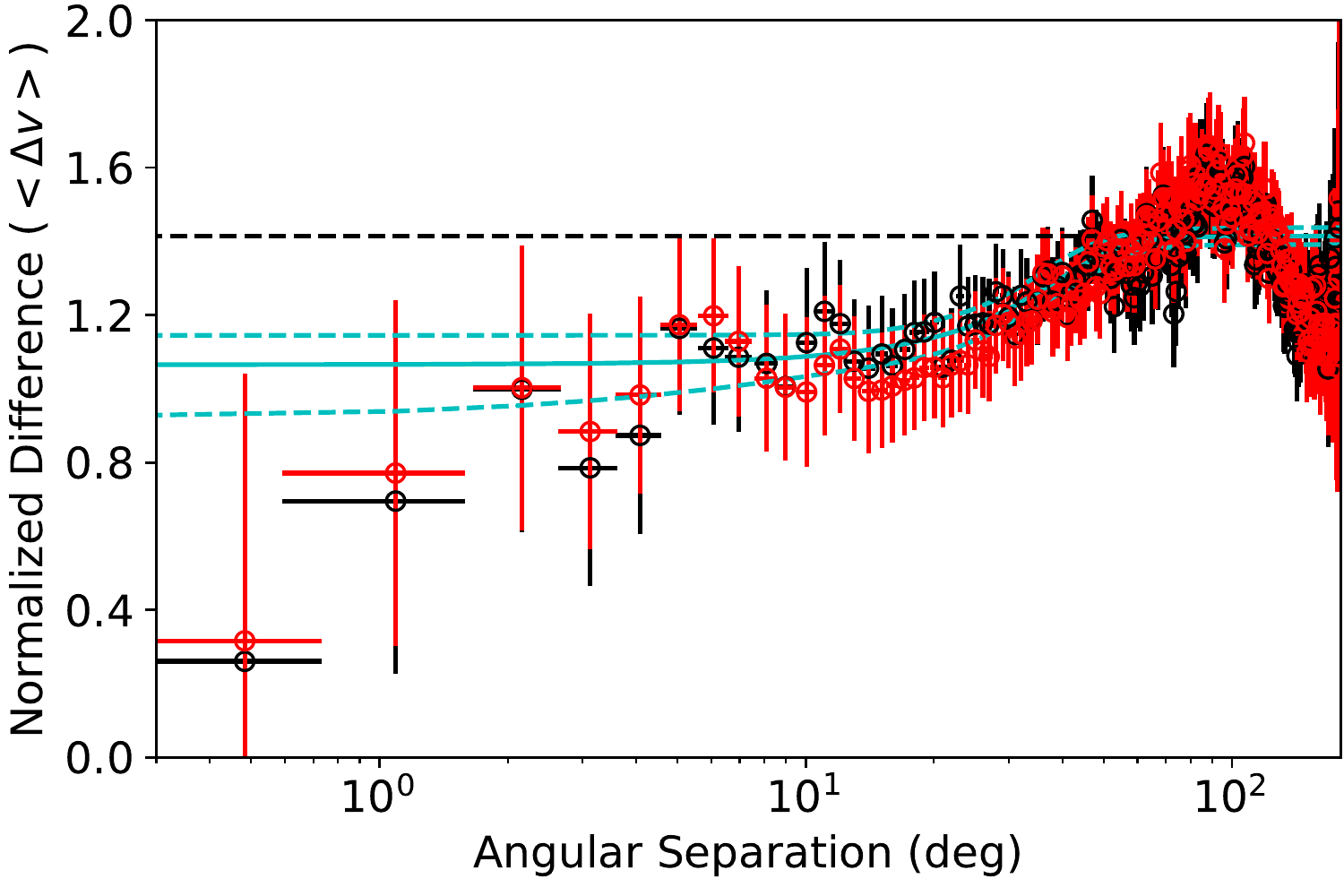}
\end{center}
\caption{Two extractions for the small scale features -- the auto-correlation function (left panels) and the structure functions (right panels) described in the Appendix.
The top panels are for the column density, while the bottom panels are for the velocity.
\ion{Si}{4} and \ion{C}{4} are shown in black and red circles, respectively, while the cyan lines are the joint models with $1\sigma$ uncertainties (dashed lines).
The extracted correlation angles are ${12.6^\circ}_{-3.6^\circ}^{+1.8^\circ}$ and $50^\circ\pm 10^\circ$ for the column density and the velocity.
}
\label{ac_res}
\end{figure*}

We consider two measurements to model the small scale features.
The first measurement is the absolute difference between sight line pairs, which is similar to the velocity structure function, and was adopted in \citetalias{Qu:2020aa}:
\begin{equation}
\Delta \log N(\theta)  = {\rm Med} (|\log N_{i} - \log N_{j}|),
\end{equation}
where $i$ and $j$ are indexes of the two sight lines with an angular distance of $\theta$.
We extract the absolute difference for both original measurements and residuals after the kinematical model correction (Fig. \ref{logN_vc_sf}).
By comparing the original and residual extraction, \citetalias{Qu:2020aa} argued that there are three angular separate regimes for the warm gas for both column density variations and kinematic structures: the random part, the global variation (e.g., rotating disk), and the cloud-size variation.
This absolute difference is also normalized so that it is dimensionless for further modeling:
\begin{equation}
\gamma (\theta) = \frac{\Delta \log N(\theta)}{A \sigma},
\end{equation}
where $\sigma$ is the standard deviation of the logarithmic column density (and the velocity).
The factor $A = 2^{1/2} {\rm erf}^{-1} (1/2)$ is the conversion factor between $\sigma$ and the median (Med) of a half-normal distribution as ${\rm Med} = A \sigma$.
When there is no correlation between the sample, the maximum of $\gamma$ is $2^{1/2}$ considering the uncertainty propagation (the right panels of Fig. \ref{ac_res}).

Another measurements is the auto-correlation coefficient, which is defined as 
\begin{equation}
 \phi (\theta)  =\frac{{\rm Med} [(\log N_{i}-\log N_{\rm md})  (\log N_{j}-\log N_{\rm md})]}{A^2 \sigma^2},
\end{equation}
where $\log N_{\rm md}$ is the median of the sample.
The factor $A^2$ normalizes the auto-correlation coefficient to 1 at zero angular separation (i.e., the standard deviation).
This auto-correlation coefficient is different from the standard correlation coefficient, where the mean is used to generate a zero-mean sample.
The median is chosen to reduce the effect of the outliers for the column density and the velocity measurements for the mean.
The two extractions for both \ion{Si}{4} and \ion{C}{4} are show in Fig. \ref{ac_res}.

Theoretically, both $\gamma(\theta)$ and $\phi(\theta)$ should capture the same features but in different ways.
In practice, a combination of these two measurements could improve the determination of the small-scale correlation, which may be affected by the measurement uncertainty and the deviation from assumptions.
In the Appendix, we introduce a statistical framework to model both the absolute difference and the auto-correlation coefficient simultaneously.

In this model, we assume that the column density or the velocity pairs follow dependent 2-dimensional Gaussian distributions with different correlation coefficients at different angular separations.
Then the absolute difference and the auto-correlation coefficient can be calculated, which is derived in the Appendix.
We fit \ion{Si}{4} and \ion{C}{4} simultaneously, because they show very similar features for both column density and velocity variations (Fig. \ref{logN_vc_sf}).
The fitting results are shown in Fig. \ref{ac_res} and Table \ref{fitting_results}.

The correlation angles for the column density and the velocity variations are ${12.6^\circ}_{-3.6^\circ}^{+1.9^\circ}$ and ${48^\circ}_{-11^\circ}^{+12^\circ}$, respectively.
These two values are roughly consistent with \citetalias{Qu:2020aa}, which suggests that there are multiple warm gas clouds within a kinematical structure (i.e., inflow and outflow).
One caveat is that the residual velocity has large-scale variations up to $90^\circ$, which breaks the assumption of the non-correlation at large angular separations.
This issue will be addressed in analyses of the difference map between \ion{Si}{4} and \ion{C}{4} (Section 4.3).

As stated in the Appendix, the ratio ($C$) is a free parameter between the measurement uncertainty over the sample standard deviation ($\sigma$), which is determined in the modeling (Appendix).
Then, we compare the derived measurement uncertainty calculated from $C$ with the observed uncertainty to check the self-consistency of the statistical framework.
The derived measurement uncertainty is $C\sigma < 0.10$ dex ($2\sigma$ upper limit) for the column density, which is consistent with the column density uncertainty distribution of individual sight lines (mostly $< 0.05$ dex).
However, this modeled uncertainty is about $C\sigma \approx 18 \kms$ for the velocity variation, which is larger than the median of the measurement uncertainty of $\approx 5\kms$.
This large derived measurement uncertainty implies another contributor of $\approx 17 \kms$ to the observed scatter at small angular separation.
This additional variation may be due to the instrumentation wavelength calibration with an uncertainty of $15-20 \kms$ \citep{Wakker:2015aa}.

\begin{table*}
\begin{center}
\caption{The Parametric Models of the Small-Scale Features}
\label{fitting_results}
\begin{tabular}{lccccccccc}
\hline
\hline
Data & $\theta_{\rm b}$ & $\alpha_1$ & $\delta$ &$\sigma$ & $C$ & $\sigma_{\rm 0, abs}^2$ & $\sigma_{\rm 0, mp}^2$\\
 & (deg) &  &  & ($\cmsq$ or $\kms$) &  &  &  \\
\hline
$\log N$ (\ion{Si}{4} and \ion{C}{4})$^a$ & $12.6_{-3.6}^{+1.9}$ & $0.47_{-0.12}^{+0.37}$ & $< 0.56$ & $0.180\pm 0.001$ & $< 0.55$ & $0.089_{-0.010}^{+0.011}$ & $1.31_{-0.14}^{+0.16}$ \\
$v_{\rm c}$ (\ion{Si}{4} and \ion{C}{4})$^a$  & $48_{-11}^{+12}$ & $1.7_{-1.0}^{+1.9}$ & $< 0.29$ & $22.7\pm 0.2$ & $0.79_{-0.07}^{+0.03}$ & $0.68_{-0.07}^{+0.08}$ & $1.98_{-0.20}^{+0.23}$ \\
$\log N$ (\ion{Si}{4} - \ion{C}{4})  & $6.1_{-4.6}^{+21.2}$ & $0.60_{-0.38}^{+1.00}$ & $0.92_{-0.31}^{+0.25}$ & $0.168\pm 0.002$ & $0.50_{-0.24}^{+0.23}$ & $0.27\pm 0.04$ & $1.25_{-0.18}^{+0.22}$ \\
$v_{\rm c}$ (\ion{Si}{4} - \ion{C}{4})  & $...$ & $...$ & $...$ & $16.3\pm 0.1$ & $0.99_{-0.21}^{+0.02}$ & $0.013\pm0.002$ & $1.22_{-0.17}^{+0.19}$ \\


\hline
\end{tabular}
\end{center}
$^a$ The joint fitting of both kinematical model-corrected \ion{C}{4} and \ion{Si}{4}.
\end{table*}

\section{The Difference between the \ion{Si}{4} and \ion{C}{4} Gases}
As shown in Section 3, the first-order kinematical models of the \ion{Si}{4} and the \ion{C}{4}-bearing gases are consistent with each other.
This consistency indicates that these gases are roughly cospatial at a large scale, which is dominated by the rotating disk of the MW warm gas.
Also, we show that there are small-scale features after subtracting the kinematical model from observed column density and velocity measurements.
In this section, we focus on the difference between the observed \ion{Si}{4} and \ion{C}{4} properties to study both the global variation and small scale features.

\subsection{Transition of the SiIV/CIV ratio - the lower Ratio GC Region}


The observed \ion{Si}{4} to \ion{C}{4} ratio ($\log N({\rm SiIV}) - \log N({\rm CIV})$) has a median of $-0.57$ dex with a scatter of 0.14 dex.
This ratio is dominated by the differences of the abundance and the ionization fraction.
Adopting the Solar abundance from \citet{Asplund:2009aa}, the silicon to carbon ratio is -0.91 dex.
The ionization ratio difference is about $+0.3$ dex in both photoionization and collisional ionization \citep{Gnat:2007aa, Oppenheimer:2013aa}.
Therefore, the predicted ion ratio is about $-0.6$ dex, which is consistent with the observed ratio.
This observed ratio is also consistent with previous studies \citep{Savage:2009aa, Wakker:2012aa}.

\begin{figure*}
\begin{center}
\includegraphics[width=0.48\textwidth]{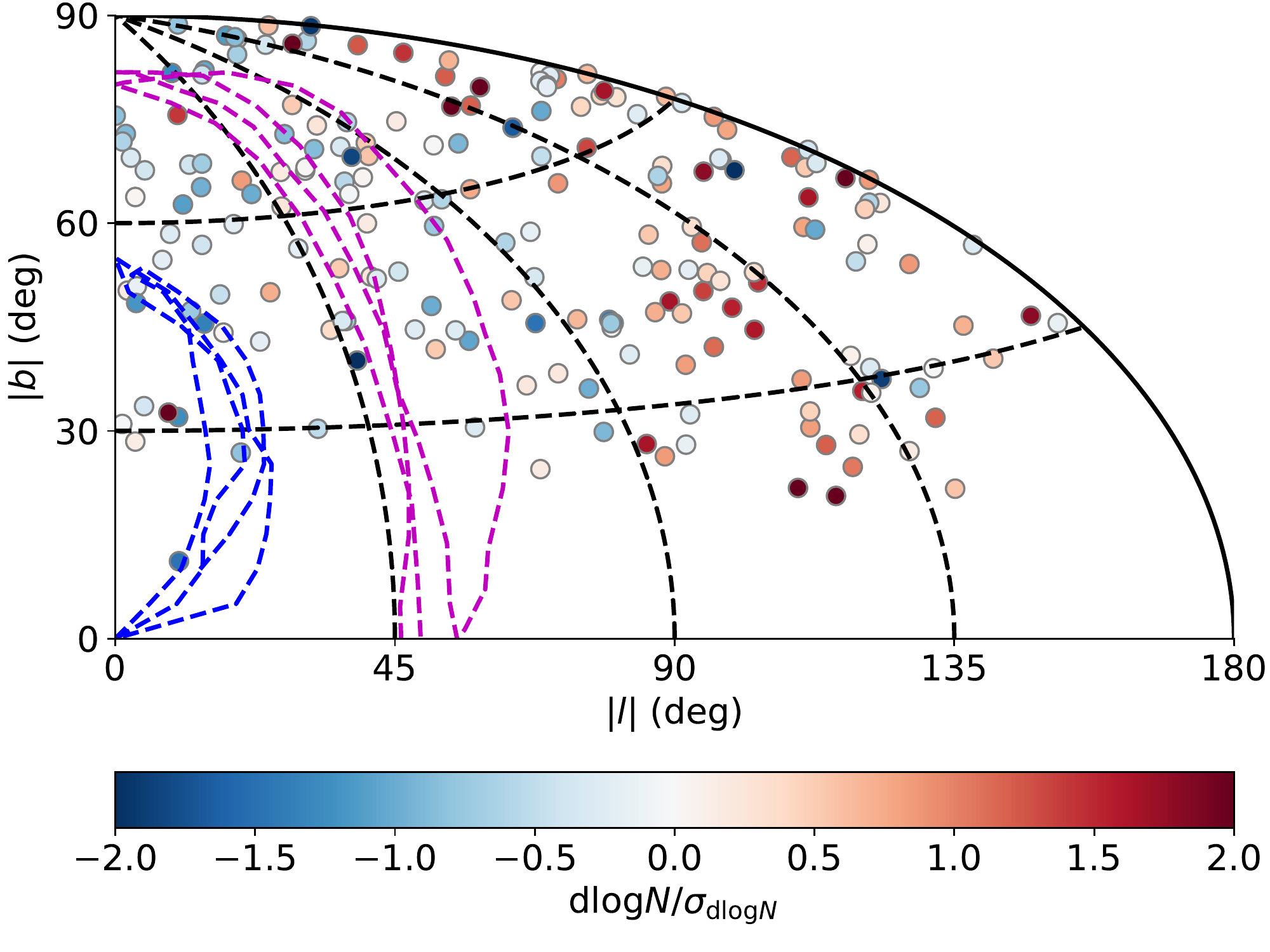}
\hspace{0.1cm}
\includegraphics[width=0.48\textwidth]{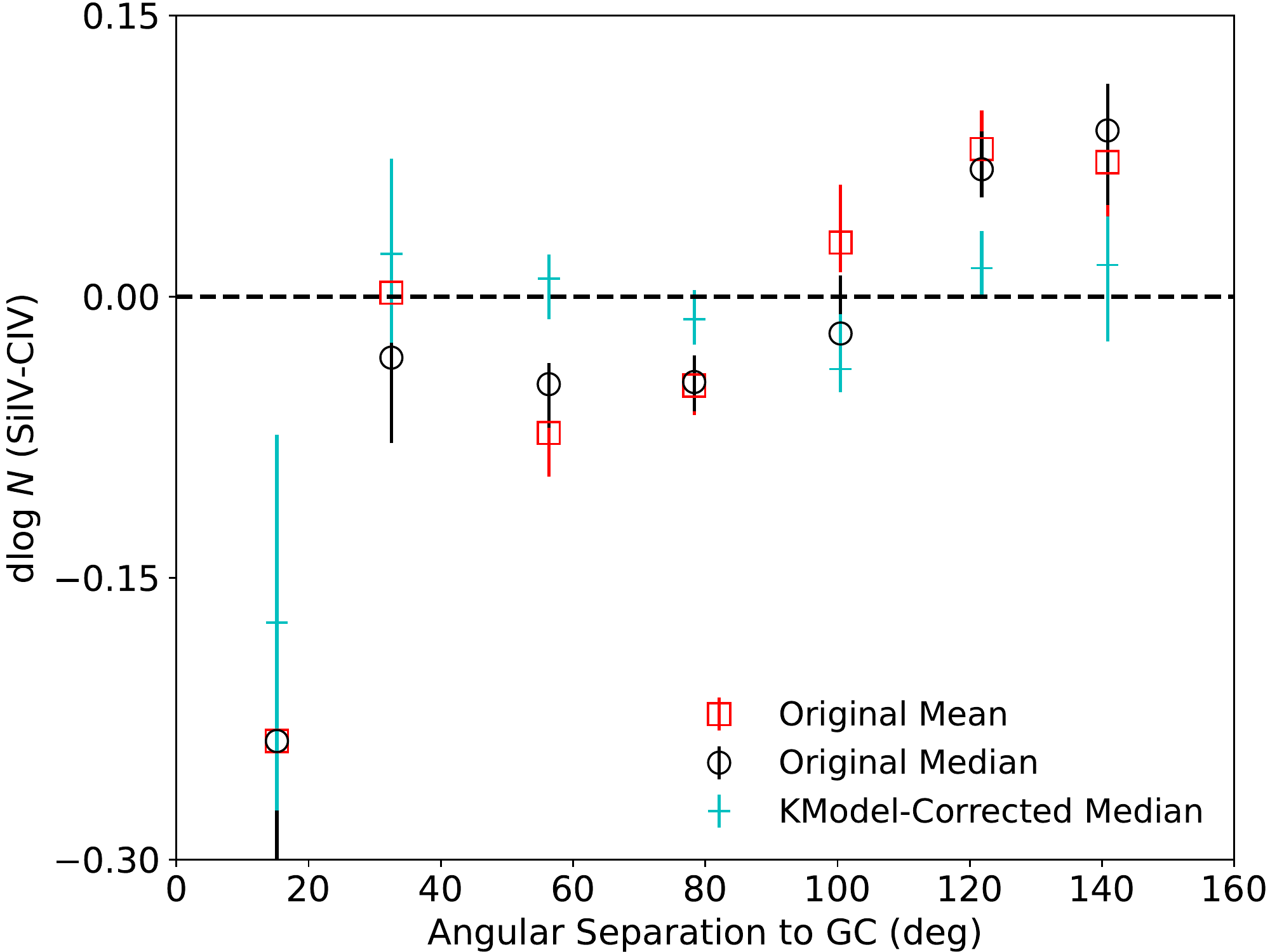}
\end{center}
\caption{The folded sky map for the \ion{Si}{4}/\ion{C}{4} column density ratio (left panel; corrected for zero-mean) and its radial profile related to the GC (right panel). 
There is a trend of increasing ion ratio from the GC to the anti-GC direction.
The boundaries of the Fermi Bubbles \citep{Su:2010aa} and the eROSITA Bubble \citep{Predehl:2020aa} are shown as blue and purple dashed lines in the left panel.
In the right panel, we also show the model corrected ratios (cyan) as a control group, which shows that after the model correction, the trend with angle has disappeared, so the kinematical model shows the global variation of the warm gas distribution.
The uncertainties in the right panel are the statistical uncertainties within each bin, except for the first bin $<20^\circ$.
There is only one sight lines at $<20^\circ$, so the uncertainty is the measurement uncertainty.}
\label{sic_dlogN}
\end{figure*}

However, we notice that the variation of this ratio is not randomly distributed over the sky.
The \ion{Si}{4}/\ion{C}{4} ratio is lower around the GC (Fig. \ref{sic_dlogN}). 
This leads to an increase of the ratio with a larger angular distance from the GC.
This trend is not reported in previous studies, mainly because of the lack of sight lines around the GC (e.g., $\lesssim 5$ sight lines are within $60^\circ$ in \citealt{Wakker:2012aa}).
The ratios averaged in annuli with increasing distance ($>6\sigma$) from the GC is also shown in Fig. \ref{sic_dlogN}, where the ratio is corrected by the median (i.e., zero median).
We also show the difference of the residual after the kinematical model correction, where the trend of the increasing ratio disappears. 
This phenomenon suggests that the gas distributions are different between the \ion{C}{4} and the \ion{Si}{4}, although parameters are consistent in the kinematical model.

We compare this low \ion{Si}{4}/\ion{C}{4} ratio region with other GC features, such as the Fermi bubble and the eROSITA bubble.
It is clear that the Fermi bubble is much smaller than the low ratio region, while the eROSITA bubble is comparable.
However, the boundary of the low \ion{Si}{4}/\ion{C}{4} ratio is not well determined because of the relative low resolution of the QSO sight line mapping to determine whether it shares a similar boundary of the eROSITA bubble.

The origin of this low \ion{Si}{4}/\ion{C}{4} ratio region is uncertain, and there are several possibilities.
First, this ion ratio difference may be a result of the variation of the element abundance.
The stellar population provides helpful insights into the Galactic scale element abundance variation, although the abundance variation of the gaseous phase is poorly known.
Spectroscopic surveys around the Solar system and GC suggest that the C/$\alpha$-element ratio decreases with higher metallicity, where $\alpha$-elements include Mg, Si, Ca, and Ti \citep[e.g.,][]{Buder:2018aa, Griffith:2019aa}.
Relative to the solar metallicity, increasing the metallicity by one order of magnitude reduces the C/$\alpha$-element abundance by about 0.3 dex.
If this empirical relationship is similar for the gaseous phase, the observed variation of the \ion{Si}{4}/\ion{C}{4} ratio ($0.2-0.3$ dex; Fig. \ref{sic_dlogN}) implies that the metallicity around the GC is about $0.5-1$ dex lower than the anti-GC direction.
Although the metallicity around the GC still has large uncertainties varying from the Solar metallicity to several times the Solar metallicity \citep{Najarro:2006aa}, it is very unlikely to have a lower metallicity than the Solar neighbor.

This ion ratio variation is also possibly due to the phase transition within the same ionization mechanism.
Although \ion{Si}{4} and \ion{C}{4} have close ionization potentials ($33 - 45$ eV and $47 - 64$ eV), they are still tracing sightly different gases.
If the warm gas is dominated by collisional ionization, the \ion{C}{4} gas has a higher temperature. 
Another possibility is that the warm gas around the GC has different ionization mechanisms.
\citet{Wakker:2012aa} compared theoretical predictions of eight mechanisms for the \ion{Si}{4}/\ion{C}{4} ratio.
These models predict a wide range of \ion{Si}{4}/\ion{C}{4} ratio from -1.5 dex to -0.3 dex, which covers the observed \ion{Si}{4}/\ion{C}{4}ratio of -1.0 to -0.5 dex.
However, \citet{Wakker:2012aa} found that none of them can reproduce all intermediate ions (i.e., \ion{Si}{4}, \ion{C}{4}, \ion{N}{5}, and \ion{O}{6}), which suggests a hybrid model is needed to explain the generation of the warm gas.

\subsection{Redshifted CIV?}
The velocity difference between the \ion{Si}{4} and \ion{C}{4} line centroids show an unexpected none-zero median of $4 \kms$ with a scatter of $16 \kms$.
Then, this non-zero median is statistically significant ($\approx4\sigma$) with 198 sight lines in the \ion{Si}{4} and \ion{C}{4} with a stacked uncertainty of $1\kms$.
This feature can be fundamental if it is a hint that the hot gas is outflowing, because the \ion{C}{4}-bearing gas traces the hotter gas compared to \ion{Si}{4}.
However, the observation significance of this feature does not mean that the non-zero shift between \ion{Si}{4} and \ion{C}{4} is physical, as it may be due to the limiations of the {\it HST}/COS.

This redshifted \ion{C}{4} feature may be due to the artifacts of the {\it HST}/COS, which is due to the geometric distortion of the instrument and the wavelength solution obtained from CALCOS. 
Empirically, this shift could lead to an uncertainty up to $20-30 \kms$ \citep[e.g.][]{Wakker:2015aa}, which may account for the additional uncertainty in the measured line centroid (Section 3.3).
Here, we summarize both pros and cons for whether this feature is physical.

Two indirect lines of evidence support this shift as being physical.
First, we consider the absorption features around the MS and M31 as another reference of the instrumental shift.
If the redshifted \ion{C}{4} is an artificial feature, it is expected to have similar shifts for both MW absorption and MS HVCs or M31 features.
However, the MS and M31 features show an opposite shift (i.e., $10 \kms$ blue shifted \ion{C}{4}), which leads to a sharp change of $\approx 15 \kms$ within a $\approx 1 \rm~ \AA$ region.
The sharp change is unlikely to be an instrumental issue, because the wavelength calibration shift is normally a smooth function over the entire spectrum ($600 \rm~ \AA$; \citealt{Wakker:2015aa}).

Second, we compare the \ion{C}{4} centroid to that of \ion{O}{6} from \citet{Savage:2003aa} and \citet{Sembach:2003aa}.
\citet{Savage:2003aa} summarized the MW major absorption features around $v\approx 0 \kms$, while \citet{Sembach:2003aa} focused on HVCs with $|v| > 90 \kms$.
For self consistency with the \ion{C}{4} and \ion{Si}{4} samples, we exclude HVCs associated with the MS and M31, and calculate the final line centroid, combining the \citet{Savage:2003aa} and \citet{Sembach:2003aa} measurements.
The joint sample of three ions has 23 sight lines.
We notice that \ion{O}{6} is also redshifted compared to \ion{Si}{4}, and even more redshifted than \ion{C}{4} with a median of $\approx 10 \kms$.
Then, there seems to be a trend that the higher the ionization state lines are more redshifted.


Previously, \citet{Wakker:2012aa} found no velocity difference, within the uncertainty, between \ion{C}{4}, \ion{Si}{4}, and \ion{O}{6} using a joint sample with tens of sight lines obtained by GHRS, STIS, and FUSE.
Compared to the {\it HST}/COS sample, these observations have lower resolutions, but do not have the wavelength calibration issue.
The \citet{Wakker:2012aa} sample has 24 sight lines with both \ion{Si}{4} and \ion{C}{4} measurements.
It is possible that the \citet{Wakker:2012aa} sample is a 24-sight line subsample with zero shift among a larger redshifted distribution (i.e., the {\it HST}/COS sample with 198 sight lines).
We test this possibility, and it is within $2 \sigma$ to have such a subsample with a zero shift.
However, we also check the sight lines ($\approx 10$) observed by both the COS or in \citet{Wakker:2012aa}, and half of them have consistent velocities between \ion{Si}{4} and \ion{C}{4}.
For another half overlapping sample, the COS \ion{C}{4} line centroid is $\approx 5-10 \kms$ redshifted compared to the STIS measurements, which suggests that some COS measurements do have biases introduced by the wavelength calibration.
However, it is not conclusive, as we are limited by the small overlap sample.
Here, we cannot determine whether the redshifted \ion{C}{4} feature seen by {\it HST}/COS is physical or artificial.

\subsection{Small-Scale Structures}
In Section 3.3, we use the kinematical model to correct the global variation and study the small scale features, showing that the cloud-size variation of $\approx 12^\circ$ and $\approx 50^\circ$ for the column density and the kinematics.
Here, another way to consider the small scale structure is using the \ion{Si}{4} - \ion{C}{4} difference maps, since these two ions are roughly cospatial at a large scale (Section 3.1).
In Fig. \ref{logN_vc_sf}, we show the structure functions of the \ion{Si}{4}/\ion{C}{4} column density ratio and the velocity difference.
Compared to the kinematical model correction, the difference maps further suppress the large-scale features.

The structure function of the column density still has a tiny peak around $90^\circ$, which accounts for the global variation described in section 4.1.
At a large scale, the behavior of the difference map is similar to the kinematical model correction (i.e., roughly random variation $> 20^\circ$).
However, at a small scale the difference map shows a smaller correlation scale ($5^\circ- 10^\circ $), while the kinematical model correction suggests a correlation angular scale of $12-15^\circ$ (Section 4.1 and \citetalias{Qu:2020aa}).

In Fig. \ref{logN_vc_sf}, we noticed that the structure function of the difference map is higher than the kinematical model correction at a small scale ($\lesssim 10^\circ$), which is due to the uncertainty propagation.
The difference map has uncertainties as a combination of two uncertainties of the \ion{Si}{4} and \ion{C}{4} measurements, so the intrinsic uncertainty will be about 1.4 times larger than the original map and the kinematical model-corrected map.
This large measurement uncertainty for every point is important when it is comparable to the structure function small scale feature.
For the column density, the measurement uncertainty is 0.05 dex, so it affects the first several bins with structure function values $\lesssim 0.1$ (angular separation $<2^\circ$).
This effect is insignificant when the structure function is higher than $0.1$.
Therefore, between $5-15^\circ$ the structure function difference between the difference map and the kinematical model correction is not due to the uncertainty propagation.

Then, we suggest that the small scale feature ($<15^\circ$) can be divided into two regimes, compared to \citetalias{Qu:2020aa}.
At a small scale ($\lesssim 5^\circ-10^\circ$), the variation is dominated by small-scale turbulence (also see the variation of the velocity described below).
Between $5^\circ-15^\circ$, there may be a cloud-scale phase transition, because the ionization state of \ion{C}{4} is slightly higher than \ion{Si}{4}.

\begin{figure*}
\begin{center}
\includegraphics[width=0.48\textwidth]{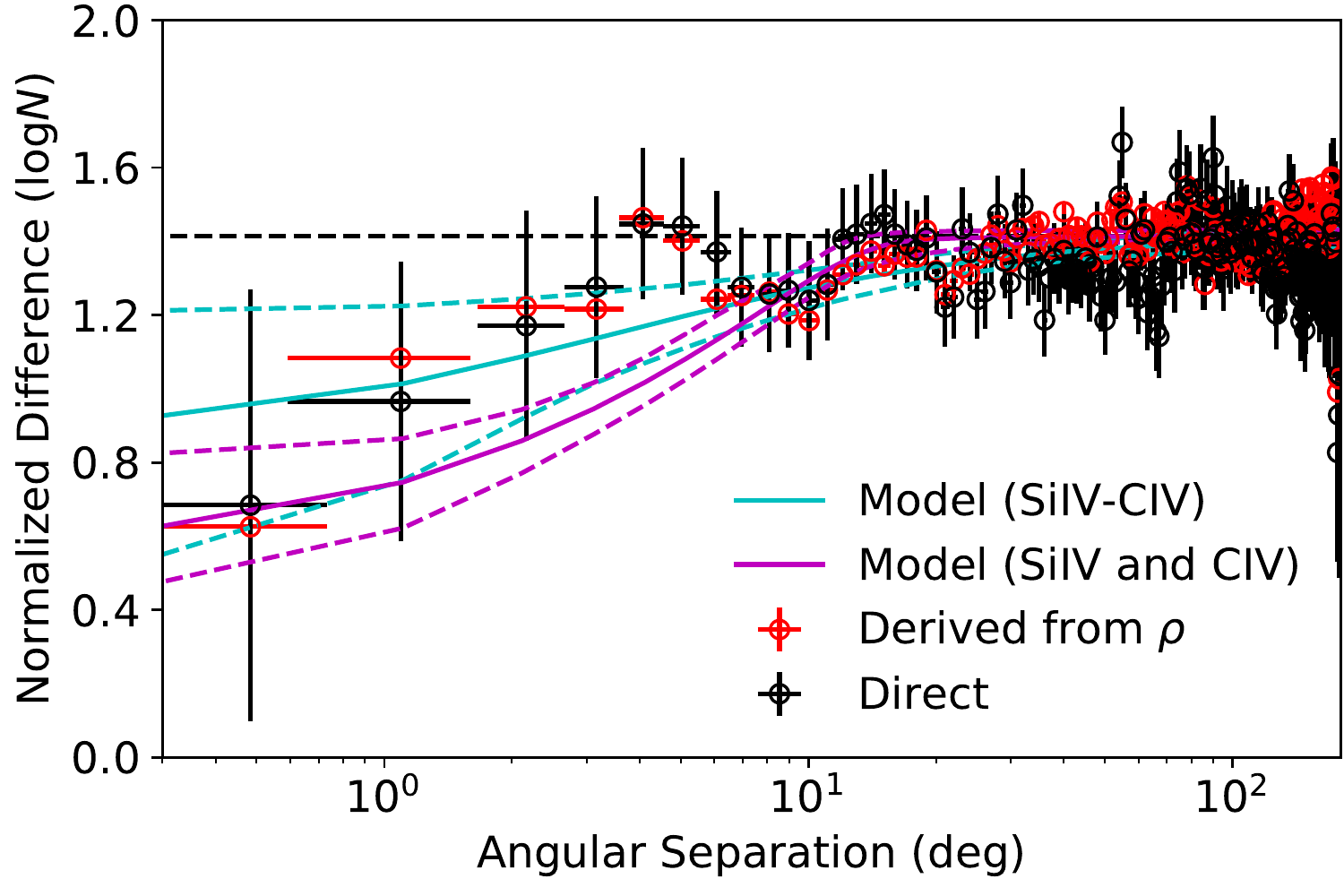}
\includegraphics[width=0.48\textwidth]{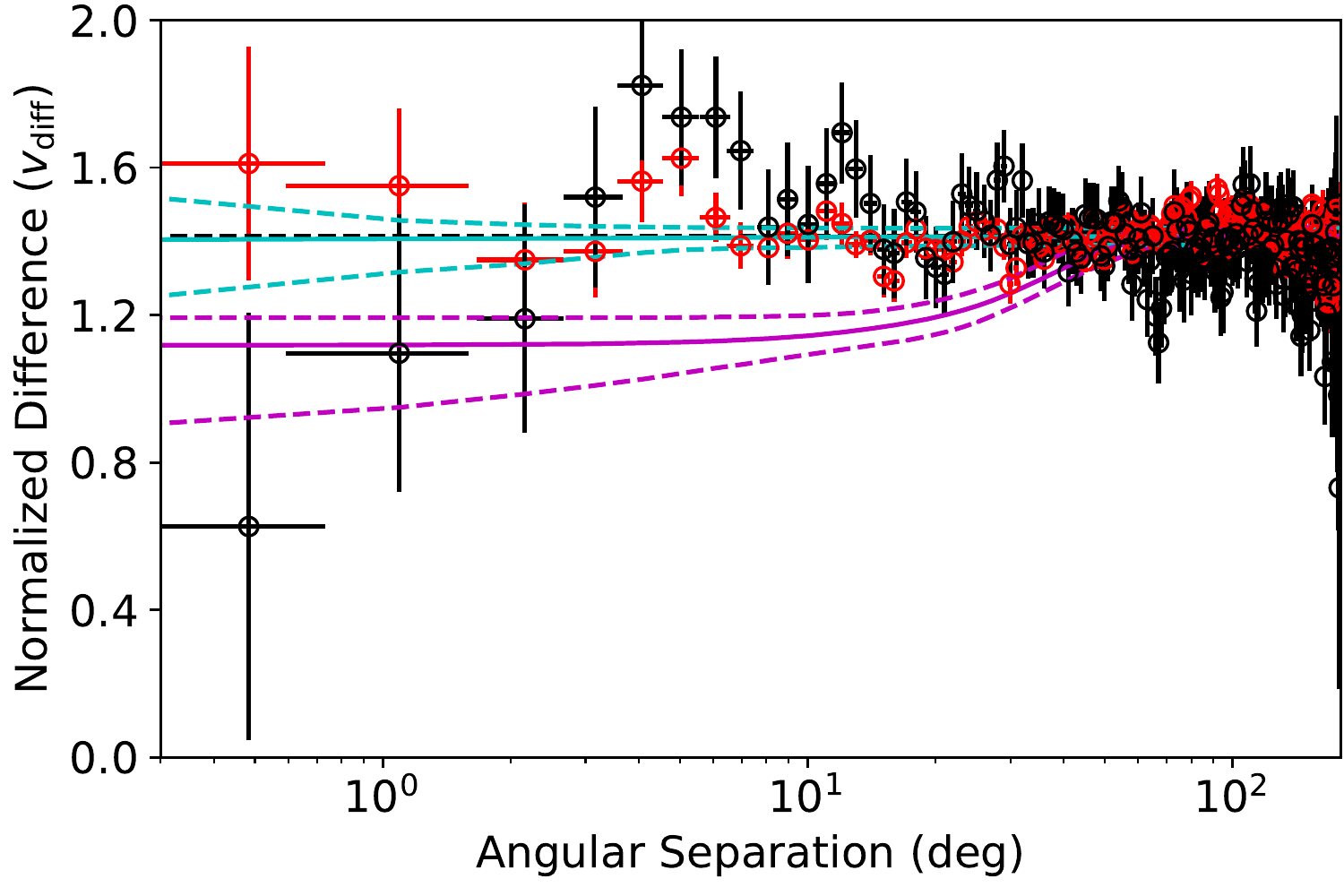}
\end{center}
\caption{The comparison of the two methods to determine the structure functions for the column density (left panel) and the velocity line centroid (right panel).
As stated in the Appendix, the structure function is defined as the mean of the absolute difference (black circles).
Under the Gaussian distribution assumption, the structure function can also be derived from the correlation coefficient (red circles).
These two methods are consistent for the density distribution and the velocity.
For the velocity, the difference at a small scale indicates that the correlation within $5^\circ$ shown in Fig. \ref{logN_vc_sf} is tentative (see the text for details).
}
\label{abd_rho}
\end{figure*}

This difference between small-scale turbulence and cloud-scale variation is more significant for the velocity structure function, when comparing the difference map extraction with the kinematical model correction.
For the kinematical model correction, there are residuals at a scale of $30^\circ-50^\circ$, which is the inflow and outflow of the Galactic fountain \citepalias{Qu:2020aa}.
However, the difference map suppresses all correlations $> 10^\circ$, which indicates that \ion{Si}{4} and \ion{C}{4}-bearing gases share the same bulk velocity field, including rotation, inflow, and outflow.
Similar to the column density, the velocity structure function is dominated by turbulence at $<10^\circ$ with a slope of 1/3 for classic Kolmogorov turbulence.

One caveat is that the relative uncertainty of the velocity ($10-20 \kms$ compared to the velocity structure function $10-15\kms$) is much larger than the column density relative uncertainty.
At a small scale, the increasing column density structure function is $>3\sigma$, while the velocity structure function tendency is $< 2\sigma$.
Therefore, the small-scale velocity structure function is still uncertain, which can be demonstrated by another approach.
As described in the Appendix, the structure function is directly related to the correlation coefficient of the sample, for a Gaussian distribution, so we can derive a structure function based on the observed correlation coefficient.
Then, there are two ways to calculate the structure function, one is the calculation of the difference between pairs, and another is derived from the correlation coefficient.
As shown in Fig. \ref{abd_rho}, these two methods are consistent for the column density.
However, the velocity structure functions are different from these two methods.
This difference may be due to uncertainty by chance ($\lesssim 2 \sigma$), or it may be because of a deviation from the Gaussian assumption.

Using the difference maps, the correlation scale is extracted using the method in Appendix, and the model parameters are listed in Table \ref{fitting_results}.
The angular size of the column density correlation scale is ${12.8^\circ}_{-8.3^\circ}^{+18.5^\circ}$, which is consistent with the joint modeling of the kinematical model-corrected \ion{Si}{4} and \ion{C}{4}.
There is no correlation for the velocity difference.
As shown in Fig. \ref{abd_rho}, the fitted model are consistent with both extractions for the velocity structure function within $1 \sigma$.

The extracted angular scale can be converted to a physical size by adopting a mean distance of 5 kpc (determined from the density distribution).
The derived physical size is 1.1 kpc, which is consistent with upper limits derived from the cloud number density of  $\approx 0.6$ per kpc around the Sun ($\approx 1.6$ kpc).
Besides the average distance, there is also a lower limit of the cloud distance, which determines the lower limits of the cloud size.
Around the Sun, there is the Local Bubble (LB) with very few \ion{Si}{4} and \ion{C}{4} clouds, which is suggested by COS WD stellar sight lines showing no absorption features \citepalias{Qu:2020aa}.
The size of the LB is about $0.2-0.5$ kpc, so the lower limit of the warm gas cloud size is about 0.1 kpc.


\section{Summary}
In this paper, we extract the all-sky MW \ion{C}{4} column density, velocity centroid and line shape from the {\it HST}/COS archive.
We analyze the density distribution and the kinematics of the \ion{C}{4}-bearing gas by adopting the kinematical model from \citetalias{Qu:2020aa}.
This new \ion{C}{4} sample is complementary to the existing \ion{Si}{4} sample (\citealt{Zheng:2019aa}; \citetalias{Qu:2020aa}), so we investigated both global and small scale variation of the \ion{Si}{4}/\ion{C}{4} ratio and kinematics.
The key results are:

\begin{itemize}

\item We develop a hybrid method to extract partially-saturated \ion{C}{4} column density line shapes by combining the AODM \citep{Savage:1991aa} and the Voigt profile fitting.
In this method, the Voigt profile fitting captures the first-order structure in the column density line shape, which accounts for saturation.
The remaining absorption column density is calculated by applying the AODM to the residuals of the Voigt profile.
The final column density line shape is the summation of both these two components.

\item There are two possible solutions to the \ion{C}{4} density distribution with similar significance in the kinematical modeling.
These two solutions can be distinguished by whether there is an extended disk component along the midplane.
Both solutions predict a similar mass of $\log M/M_\odot = 7.93\pm 0.05$ for the warm gas disk, and an upper limit of $\log M/M_\odot < 9.3$ for the CGM component.

\item Our modeling suggests that the two possible solutions of \ion{C}{4} density distribution and corresponding kinematics are consistent with the \ion{Si}{4} -- cospatial at large scale ($>10^\circ$) and sharing similar bulk velocity fields (i.e., Galactic rotation and inflow in the north).
However,  there are still large-scale variations for the \ion{Si}{4}/\ion{C}{4} ion column density ratios.
In Fig. \ref{sic_dlogN}, we show this ratio is lower around the GC, which may indicate a global phase transition of the warm gas in the MW (e.g., high temperature or harder ionizing field around the GC).

\item The difference map of the line centroids shows an unexpected redshift for \ion{C}{4} compared to the \ion{Si}{4} velocity.
The reason for this redshift is still unknown using the {\it HST}/COS sample.
This feature may be due to the artificial shift of the instrument calibration.
{\it HST}/COS has a well-known instrumental uncertainty of velocity measurement with a scatter of $15-20 \kms$, but it is not clear whether it induces systematic shifts between \ion{Si}{4} and \ion{C}{4}.
On the other hand, HVCs around M31 and LMC show a blueshifted \ion{C}{4}, which is opposite to the shift in the MW absorption features.
Different shifted velocities suggest that these shifts may be physical instead of artificial, since it is unlikely for the instrumental shift to change within a narrow wavelength range.

\item We adopted two methods to characterize the small-scale features -- the kinematical model correction and the \ion{Si}{4}-\ion{C}{4} difference map.
The results from the kinematical model correction are consistent with \citetalias{Qu:2020aa}, showing $\approx 15^\circ$ and $\approx 30^\circ-50^\circ$ residual features, which are the cloud-size and inflow or outflow scale.
The difference map can be used to study the small-scale features because \ion{Si}{4} and \ion{C}{4} are roughly cospatial at a large scale.
Comparing to the model-corrected results, the difference map can further suppress the large-scale features.
For both the column density and velocity, features $\gtrsim 10^\circ$ are suppressed, suggesting a small correlation scale ($\lesssim 10^\circ$; Fig. \ref{logN_vc_sf}).
This smallest regime may be dominated by turbulence within individual clouds.

\end{itemize}

\begin{acknowledgments}
The authors would like to thank the anonymous referee, Jiang-Tao Li, Yuan Li, and Yong Zheng for kindly helps and thoughtful discussions on this work.
Z.Q. acknowledges Astropy \citep{Astropy-Collaboration:2013aa}, Emcee \citep{Foreman-Mackey:2013aa}, and HSLA \citep{Peeples:2017aa} to making various resources public, without which this work might be delayed by years. 
This work is supported by NASA through a grant HST-AR-15806.002-A from the Space Telescope Science Institute.
\end{acknowledgments}

\bibliography{MissingBaryon}{}

\begin{thebibliography}{}
\expandafter\ifx\csname natexlab\endcsname\relax\def\natexlab#1{#1}\fi

\bibitem[{{Asplund} {et~al.}(2009){Asplund}, {Grevesse}, {Sauval}, \&
  {Scott}}]{Asplund:2009aa}
{Asplund}, M., {Grevesse}, N., {Sauval}, A.~J., \& {Scott}, P. 2009, \araa, 47,
  481

\bibitem[{{Astropy Collaboration} {et~al.}(2013){Astropy Collaboration},
  {Robitaille}, {Tollerud}, {Greenfield}, {Droettboom}, {Bray}, {Aldcroft},
  {Davis}, {Ginsburg}, {Price-Whelan}, {Kerzendorf}, {Conley}, {Crighton},
  {Barbary}, {Muna}, {Ferguson}, {Grollier}, {Parikh}, {Nair}, {Unther},
  {Deil}, {Woillez}, {Conseil}, {Kramer}, {Turner}, {Singer}, {Fox}, {Weaver},
  {Zabalza}, {Edwards}, {Azalee Bostroem}, {Burke}, {Casey}, {Crawford},
  {Dencheva}, {Ely}, {Jenness}, {Labrie}, {Lim}, {Pierfederici}, {Pontzen},
  {Ptak}, {Refsdal}, {Servillat}, \&
  {Streicher}}]{Astropy-Collaboration:2013aa}
{Astropy Collaboration}, {Robitaille}, T.~P., {Tollerud}, E.~J., {et~al.} 2013,
  \aap, 558, A33

\bibitem[{{Bish} {et~al.}(2021){Bish}, {Werk}, {Peek}, {Zheng}, \&
  {Putman}}]{Bish:2021aa}
{Bish}, H.~V., {Werk}, J.~K., {Peek}, J., {Zheng}, Y., \& {Putman}, M. 2021,
  \apj, 912, 8

\bibitem[{{Bogd{\'a}n} {et~al.}(2017){Bogd{\'a}n}, {Bourdin}, {Forman},
  {Kraft}, {Vogelsberger}, {Hernquist}, \& {Springel}}]{Bogdan:2017aa}
{Bogd{\'a}n}, {\'A}., {Bourdin}, H., {Forman}, W.~R., {et~al.} 2017, \apj, 850,
  98

\bibitem[{{Bowen} {et~al.}(2008){Bowen}, {Jenkins}, {Tripp}, {Sembach},
  {Savage}, {Moos}, {Oegerle}, {Friedman}, {Gry}, {Kruk}, {Murphy}, {Sankrit},
  {Shull}, {Sonneborn}, \& {York}}]{Bowen:2008aa}
{Bowen}, D.~V., {Jenkins}, E.~B., {Tripp}, T.~M., {et~al.} 2008, \apjs, 176, 59

\bibitem[{{Bregman} {et~al.}(2018){Bregman}, {Anderson}, {Miller},
  {Hodges-Kluck}, {Dai}, {Li}, {Li}, \& {Qu}}]{Bregman:2018aa}
{Bregman}, J.~N., {Anderson}, M.~E., {Miller}, M.~J., {et~al.} 2018, \apj, 862,
  3

\bibitem[{{Bregman} {et~al.}(2021){Bregman}, {Hodges-Kluck}, {Qu}, {Pratt},
  {Li}, \& {Yun}}]{Bregman:2021aa}
{Bregman}, J.~N., {Hodges-Kluck}, E., {Qu}, Z., {et~al.} 2021, arXiv e-prints,
  arXiv:2107.14281

\bibitem[{{Buder} {et~al.}(2018){Buder}, {Asplund}, {Duong}, {Kos}, {Lind},
  {Ness}, {Sharma}, {Bland-Hawthorn}, {Casey}, {de Silva}, {D'Orazi},
  {Freeman}, {Lewis}, {Lin}, {Martell}, {Schlesinger}, {Simpson}, {Zucker},
  {Zwitter}, {Amarsi}, {Anguiano}, {Carollo}, {Casagrande}, {{\v{C}}otar},
  {Cottrell}, {da Costa}, {Gao}, {Hayden}, {Horner}, {Ireland}, {Kafle},
  {Munari}, {Nataf}, {Nordlander}, {Stello}, {Ting}, {Traven}, {Watson},
  {Wittenmyer}, {Wyse}, {Yong}, {Zinn}, {{\v{Z}}erjal}, \& {Galah
  Collaboration}}]{Buder:2018aa}
{Buder}, S., {Asplund}, M., {Duong}, L., {et~al.} 2018, \mnras, 478, 4513

\bibitem[{{Burchett} {et~al.}(2019){Burchett}, {Tripp}, {Prochaska}, {Werk},
  {Tumlinson}, {Howk}, {Willmer}, {Lehner}, {Meiring}, {Bowen}, {Bordoloi},
  {Peeples}, {Jenkins}, {O'Meara}, {Tejos}, \& {Katz}}]{Burchett:2019aa}
{Burchett}, J.~N., {Tripp}, T.~M., {Prochaska}, J.~X., {et~al.} 2019, \apjl,
  877, L20

\bibitem[{{Das} {et~al.}(2019){Das}, {Mathur}, {Nicastro}, \&
  {Krongold}}]{Das:2019aa}
{Das}, S., {Mathur}, S., {Nicastro}, F., \& {Krongold}, Y. 2019, \apjl, 882,
  L23

\bibitem[{{de Boer} \& {Savage}(1983)}]{de-Boer:1983aa}
{de Boer}, K.~S., \& {Savage}, B.~D. 1983, \apj, 265, 210

\bibitem[{{Fang} {et~al.}(2015){Fang}, {Buote}, {Bullock}, \&
  {Ma}}]{Fang:2015aa}
{Fang}, T., {Buote}, D., {Bullock}, J., \& {Ma}, R. 2015, \apjs, 217, 21

\bibitem[{{Foreman-Mackey} {et~al.}(2013){Foreman-Mackey}, {Hogg}, {Lang}, \&
  {Goodman}}]{Foreman-Mackey:2013aa}
{Foreman-Mackey}, D., {Hogg}, D.~W., {Lang}, D., \& {Goodman}, J. 2013, \pasp,
  125, 306

\bibitem[{{Fox} {et~al.}(2019){Fox}, {Richter}, {Ashley}, {Heckman}, {Lehner},
  {Werk}, {Bordoloi}, \& {Peeples}}]{Fox:2019aa}
{Fox}, A.~J., {Richter}, P., {Ashley}, T., {et~al.} 2019, \apj, 884, 53

\bibitem[{{Fox} {et~al.}(2004){Fox}, {Savage}, {Wakker}, {Richter}, {Sembach},
  \& {Tripp}}]{Fox:2004aa}
{Fox}, A.~J., {Savage}, B.~D., {Wakker}, B.~P., {et~al.} 2004, \apj, 602, 738

\bibitem[{{Fox} {et~al.}(2014){Fox}, {Wakker}, {Barger}, {Hernandez},
  {Richter}, {Lehner}, {Bland-Hawthorn}, {Charlton}, {Westmeier}, {Thom},
  {Tumlinson}, {Misawa}, {Howk}, {Haffner}, {Ely}, {Rodriguez-Hidalgo}, \&
  {Kumari}}]{Fox:2014aa}
{Fox}, A.~J., {Wakker}, B.~P., {Barger}, K.~A., {et~al.} 2014, \apj, 787, 147

\bibitem[{{Gnat}(2017)}]{Gnat:2017aa}
{Gnat}, O. 2017, \apjs, 228, 11

\bibitem[{{Gnat} \& {Sternberg}(2007)}]{Gnat:2007aa}
{Gnat}, O., \& {Sternberg}, A. 2007, \apjs, 168, 213

\bibitem[{{Griffith} {et~al.}(2019){Griffith}, {Johnson}, \&
  {Weinberg}}]{Griffith:2019aa}
{Griffith}, E., {Johnson}, J.~A., \& {Weinberg}, D.~H. 2019, \apj, 886, 84

\bibitem[{{Jenkins}(1978)}]{Jenkins:1978aa}
{Jenkins}, E.~B. 1978, \apj, 219, 845

\bibitem[{{Jenkins}(1996)}]{Jenkins:1996aa}
---. 1996, \apj, 471, 292

\bibitem[{{Kaaret} {et~al.}(2020){Kaaret}, {Koutroumpa}, {Kuntz}, {Jahoda},
  {Bluem}, {Gulick}, {Hodges-Kluck}, {LaRocca}, {Ringuette}, \&
  {Zajczyk}}]{Kaaret:2020aa}
{Kaaret}, P., {Koutroumpa}, D., {Kuntz}, K.~D., {et~al.} 2020, Nature
  Astronomy, 4, 1072

\bibitem[{{Kim} \& {Ostriker}(2018)}]{Kim:2018aa}
{Kim}, C.-G., \& {Ostriker}, E. 2018, \apj, 853, 173

\bibitem[{{Lehner} \& {Howk}(2011)}]{Lehner:2011aa}
{Lehner}, N., \& {Howk}, J.~C. 2011, Science, 334, 955

\bibitem[{{Lehner} {et~al.}(2020){Lehner}, {Berek}, {Howk}, {Wakker},
  {Tumlinson}, {Jenkins}, {Prochaska}, {Augustin}, {Ji}, {Faucher-Gigu{\`e}re},
  {Hafen}, {Peeples}, {Barger}, {Berg}, {Bordoloi}, {Brown}, {Fox}, {Gilbert},
  {Guhathakurta}, {Kalirai}, {Lockman}, {O'Meara}, {Pisano}, {Ribaudo}, \&
  {Werk}}]{Lehner:2020aa}
{Lehner}, N., {Berek}, S.~C., {Howk}, J.~C., {et~al.} 2020, \apj, 900, 9

\bibitem[{{Li} \& {Wang}(2013)}]{Li:2013aa}
{Li}, J.-T., \& {Wang}, Q.~D. 2013, \mnras, 428, 2085

\bibitem[{{Li} \& {Tonnesen}(2020)}]{Li:2020aa}
{Li}, M., \& {Tonnesen}, S. 2020, \apj, 898, 148

\bibitem[{{M{\'e}nard} {et~al.}(2010){M{\'e}nard}, {Scranton}, {Fukugita}, \&
  {Richards}}]{Menard:2010aa}
{M{\'e}nard}, B., {Scranton}, R., {Fukugita}, M., \& {Richards}, G. 2010,
  \mnras, 405, 1025

\bibitem[{{Miller} \& {Bregman}(2015)}]{Miller:2015aa}
{Miller}, M., \& {Bregman}, J. 2015, \apj, 800, 14

\bibitem[{Nadarajah \& Pog{\'a}ny(2016)}]{Nadarajah:2016aa}
Nadarajah, S., \& Pog{\'a}ny, T.~K. 2016, Comptes Rendus Mathematique, 354, 201

\bibitem[{{Najarro}(2006)}]{Najarro:2006aa}
{Najarro}, F. 2006, in Journal of Physics Conference Series, Vol.~54, Journal
  of Physics Conference Series, 224--232

\bibitem[{{Oppenheimer} \& {Schaye}(2013)}]{Oppenheimer:2013aa}
{Oppenheimer}, B.~D., \& {Schaye}, J. 2013, \mnras, 434, 1043

\bibitem[{{Peek} {et~al.}(2015){Peek}, {M{\'e}nard}, \&
  {Corrales}}]{Peek:2015aa}
{Peek}, J.~E.~G., {M{\'e}nard}, B., \& {Corrales}, L. 2015, \apj, 813, 7

\bibitem[{{Peeples} {et~al.}(2017){Peeples}, {Tumlinson}, {Fox}, {Aloisi},
  {Fleming}, {Jedrzejewski}, {Oliveira}, {Ayres}, {Danforth}, {Keeney}, \&
  {Jenkins}}]{Peeples:2017aa}
{Peeples}, M., {Tumlinson}, J., {Fox}, A., {et~al.} 2017, {The HSLA}, Tech.
  rep., STScI

\bibitem[{{Predehl} {et~al.}(2020){Predehl}, {Sunyaev}, {Becker}, {Brunner},
  {Burenin}, {Bykov}, {Cherepashchuk}, {Chugai}, {Churazov}, {Doroshenko},
  {Eismont}, {Freyberg}, {Gilfanov}, {Haberl}, {Khabibullin}, {Krivonos},
  {Maitra}, {Medvedev}, {Merloni}, {Nandra}, {Nazarov}, {Pavlinsky}, {Ponti},
  {Sanders}, {Sasaki}, {Sazonov}, {Strong}, \& {Wilms}}]{Predehl:2020aa}
{Predehl}, P., {Sunyaev}, R.~A., {Becker}, W., {et~al.} 2020, \nat, 588, 227

\bibitem[{{Prochaska} {et~al.}(2017){Prochaska}, {Werk}, {Worseck}, {Tripp},
  {Tumlinson}, {Burchett}, {Fox}, {Fumagalli}, {Lehner}, {Peeples}, \&
  {Tejos}}]{Prochaska:2017aa}
{Prochaska}, J.~X., {Werk}, J.~K., {Worseck}, G., {et~al.} 2017, \apj, 837, 169

\bibitem[{{Qu} {et~al.}(2020){Qu}, {Bregman}, {Hodges-Kluck}, {Li}, \&
  {Lindley}}]{Qu:2020aa}
{Qu}, Z., {Bregman}, J.~N., {Hodges-Kluck}, E., {Li}, J.-T., \& {Lindley}, R.
  2020, \apj, 894, 142

\bibitem[{{Qu} {et~al.}(2019){Qu}, {Bregman}, \& {Hodges-Kluck}}]{Qu:2019aa}
{Qu}, Z., {Bregman}, J.~N., \& {Hodges-Kluck}, E.~J. 2019, \apj, 876, 101

\bibitem[{{Savage} \& {de Boer}(1981)}]{Savage:1981aa}
{Savage}, B.~D., \& {de Boer}, K.~S. 1981, \apj, 243, 460

\bibitem[{{Savage} \& {Sembach}(1991)}]{Savage:1991aa}
{Savage}, B.~D., \& {Sembach}, K.~R. 1991, \apj, 379, 245

\bibitem[{{Savage} \& {Wakker}(2009)}]{Savage:2009aa}
{Savage}, B.~D., \& {Wakker}, B.~P. 2009, \apj, 702, 1472

\bibitem[{{Savage} {et~al.}(2003){Savage}, {Sembach}, {Wakker}, {Richter},
  {Meade}, {Jenkins}, {Shull}, {Moos}, \& {Sonneborn}}]{Savage:2003aa}
{Savage}, B.~D., {Sembach}, K.~R., {Wakker}, B.~P., {et~al.} 2003, \apjs, 146,
  125

\bibitem[{{Sembach} \& {Savage}(1992)}]{Sembach:1992aa}
{Sembach}, K.~R., \& {Savage}, B.~D. 1992, \apjs, 83, 147

\bibitem[{{Sembach} {et~al.}(2003){Sembach}, {Wakker}, {Savage}, {Richter},
  {Meade}, {Shull}, {Jenkins}, {Sonneborn}, \& {Moos}}]{Sembach:2003aa}
{Sembach}, K.~R., {Wakker}, B.~P., {Savage}, B.~D., {et~al.} 2003, \apjs, 146,
  165

\bibitem[{{Shull} \& {Slavin}(1994)}]{Shull:1994aa}
{Shull}, J.~M., \& {Slavin}, J.~D. 1994, \apj, 427, 784

\bibitem[{{Stern} {et~al.}(2018){Stern}, {Faucher-Gigu{\`e}re}, {Hennawi},
  {Hafen}, {Johnson}, \& {Fielding}}]{Stern:2018aa}
{Stern}, J., {Faucher-Gigu{\`e}re}, C.-A., {Hennawi}, J.~F., {et~al.} 2018,
  \apj, 865, 91

\bibitem[{{Stocke} {et~al.}(2013){Stocke}, {Keeney}, {Danforth}, {Shull},
  {Froning}, {Green}, {Penton}, \& {Savage}}]{Stocke:2013aa}
{Stocke}, J.~T., {Keeney}, B.~A., {Danforth}, C.~W., {et~al.} 2013, \apj, 763,
  148

\bibitem[{{Su} {et~al.}(2010){Su}, {Slatyer}, \& {Finkbeiner}}]{Su:2010aa}
{Su}, M., {Slatyer}, T.~R., \& {Finkbeiner}, D.~P. 2010, \apj, 724, 1044

\bibitem[{{Tumlinson} {et~al.}(2017){Tumlinson}, {Peeples}, \&
  {Werk}}]{Tumlinson:2017aa}
{Tumlinson}, J., {Peeples}, M.~S., \& {Werk}, J.~K. 2017, \araa, 55, 389

\bibitem[{{Wakker} {et~al.}(2015){Wakker}, {Hernandez}, {French}, {Kim},
  {Oppenheimer}, \& {Savage}}]{Wakker:2015aa}
{Wakker}, B.~P., {Hernandez}, A.~K., {French}, D.~M., {et~al.} 2015, \apj, 814,
  40

\bibitem[{{Wakker} {et~al.}(2012){Wakker}, {Savage}, {Fox}, {Benjamin}, \&
  {Shapiro}}]{Wakker:2012aa}
{Wakker}, B.~P., {Savage}, B.~D., {Fox}, A.~J., {Benjamin}, R.~A., \&
  {Shapiro}, P.~R. 2012, \apj, 749, 157

\bibitem[{{Werk} {et~al.}(2014){Werk}, {Prochaska}, {Tumlinson}, {Peeples},
  {Tripp}, {Fox}, {Lehner}, {Thom}, {O'Meara}, {Ford}, {Bordoloi}, {Katz},
  {Tejos}, {Oppenheimer}, {Dav{\'e}}, \& {Weinberg}}]{Werk:2014aa}
{Werk}, J.~K., {Prochaska}, J.~X., {Tumlinson}, J., {et~al.} 2014, \apj, 792, 8

\bibitem[{{Werk} {et~al.}(2019){Werk}, {Rubin}, {Bish}, {Prochaska}, {Zheng},
  {O{\textquoteright}Meara}, {Lenz}, {Hummels}, \& {Deason}}]{Werk:2019aa}
{Werk}, J.~K., {Rubin}, K.~H.~R., {Bish}, H.~V., {et~al.} 2019, \apj, 887, 89

\bibitem[{{Zheng} {et~al.}(2019){Zheng}, {Peek}, {Putman}, \&
  {Werk}}]{Zheng:2019aa}
{Zheng}, Y., {Peek}, J.~E.~G., {Putman}, M.~E., \& {Werk}, J.~K. 2019, \apj,
  871, 35

\bibitem[{{Zheng} {et~al.}(2017){Zheng}, {Peek}, {Werk}, \&
  {Putman}}]{Zheng:2017aa}
{Zheng}, Y., {Peek}, J.~E.~G., {Werk}, J.~K., \& {Putman}, M.~E. 2017, \apj,
  834, 179

\end{thebibliography}
\bibliographystyle{aasjournal}

\appendix

We developed a statistical framework to model the observed structure function and the auto-correlation function.
In this framework, we assume that the all-sky distributions of the column density (in logarithmic scale) and the velocity can be approximated as Gaussian distributions.
Then, the sight-line pairs at different angular separations are correlated Gaussian distributions, where the correlation coefficient ($\rho(\theta)$) varies at different angular separations.

Once $\rho(\theta)$ is known, the absolute difference functions can be calculated according to the summation of correlated Gaussian variables:
\begin{equation}
\gamma (\theta)  = \sqrt{2-2\rho(\theta)}
\end{equation}
Therefore, a positive correlation coefficient $\rho$ leads to a small median absolute difference, as shown in Fig. \ref{ac_res}.

To reproduce the auto-correlation coefficient, we adopt the product distribution of the correlated Gaussian distribution from \citet{Nadarajah:2016aa}:
\begin{equation}
p(z) = \frac{1}{\pi \sqrt{1-\rho^2}} \exp (\frac{\rho z}{1-\rho^2}) K_0 (\frac{|z|}{1-\rho^2}),
\end{equation}
where $z$ is the product of two variables normalized by $\sigma^2$, and $K_0$ is the modified Bessel function.
The auto-correlation function $\phi(\theta)$ is extracted from the $p(z)$ distribution numerically, because there is no analytic solution for the median of the product of the correlated Gaussian distribution.

Then, we need to define a parametric form for the correlation coefficient ($\rho(\theta)$), and we assumed it could be represented as a broken power law.
This is because the classical turbulence theory predicts that the velocity structure function (i.e., the absolute difference measurement) follows a power law at a small scale and is roughly flat at a large scale.
In Fig. \ref{logN_vc_sf}, both the column density and the velocity show similar power laws with slopes of $\approx 1/3$ to $1/2$ at small scales of $< 5^\circ$, which satisfy the theoretical prediction.

A smoothly broken power law with a flat part is defined as 
\begin{equation}
f(\theta) = B \left(\frac{\theta}{\theta_{\rm b}} \right)^{\alpha_1} \left[1+\left(\frac{\theta}{\theta_{\rm b}}\right)^{1/\delta} \right]^{-\alpha_1 \delta},
\end{equation}
where $B$ is the factor to normalize this function to 1 at large angular separation, $\theta_{\rm b}$ is the transition angle, and $\alpha_1$ is the slope of the power law at small angular separation.
The parameter $\delta$ is the smoothness parameter, and larger $\delta$ values indicate a smoother transition from the power law to the flat part.
Therefore, this broken power law is a function between 0 and 1.
In practice, the absolute difference cannot be zero at small separation because of the measurement uncertainty.
The final absolute difference function is
\begin{equation}
g(\theta) = \sqrt{(1-C^2) f^2(\theta) + C^2},
\end{equation}
where $C$ is the ratio of the measurement uncertainty over the standard deviation $\sigma$, which is between 0 and 1.
The, the correlation coefficient has a format of
\begin{equation}
\rho(\theta) = 1 - g^2(\theta).
\end{equation}

This model is fitted in the Bayesian inference with the likelihood 
\begin{eqnarray}
p_{\rm tot} &=& p_{\rm abs} + p_{\rm mp}, \notag \\
p_{\rm abs} &=& - \frac{1}{2} \sum \left[ \frac{(\gamma-\gamma_{\rm m})^2}{\sigma^2_{\rm abs}}  + \ln \sigma^2_{\rm abs} \right], \notag \\
p_{\rm mp} &=& - \frac{1}{2} \sum \left[ \frac{(\phi-\phi_{\rm m})^2}{\sigma^2_{\rm mp}} + \ln \sigma^2_{\rm mp} \right],
\end{eqnarray}
where $p_{\rm abs}$ and $p_{\rm mp}$ are the likelihoods for the absolute difference and the auto-correlation coefficient, while $\gamma_{\rm m}$ and $\phi_{\rm m}$ are the model predictions.
$\sigma^2_{\rm abs }$ and $\sigma^2_{\rm mp}$ are the empirical variance in the modeling, which depends on the number of pairs in each angular separation bin ($N_{\rm pairs}$).
This dependence is $\sigma^2_{\rm abs} (\theta) = \sigma^2_{\rm 0, abs }/ N_{\rm pairs}(\theta) g^2(\theta)$ for the absolute difference function, and $\sigma^2_{\rm mp} (\theta) = \sigma^2_{\rm 0, mp }/ N_{\rm pairs}(\theta)$ for the auto-correlation function.

As a summary, this model has seven free parameters: the equivalent standard deviation ($\sigma$), the break angle ($\theta_{\rm b}$), the slope of the power-law ($\alpha_1$), the smooth parameter of the broken power-law ($\delta$), the ratio between the measurement uncertainty to the standard deviation ($C$), the empirical variance of the absolute difference function ($\sigma_{\rm 0, abs}^2$), and the empirical variance of the auto-correlation function ($\sigma_{\rm 0, mp}^2$).

\end{document}